\documentclass[10pt,twocolumn,twoside]{IEEEtran}
\usepackage{cleveref}
\usepackage[dvips]{graphicx}
\usepackage{mathdots}
\usepackage{amssymb}
\usepackage{amsthm}
\usepackage{mathtools}
\usepackage{bm}
\usepackage{cite}
\usepackage{epstopdf}
\usepackage{algorithm}               
\usepackage{algorithmic}             
\usepackage{multirow}                
\usepackage{amsmath}
\usepackage{color,xcolor}
\usepackage{soul}
\usepackage{accents}
\usepackage{bigints}
\graphicspath{{../}}
\usepackage{wrapfig}
\usepackage{csvsimple}
\usepackage{colortbl}
\DeclareMathOperator*{\argmin}{arg\,min} 
\DeclareMathOperator*{\argmax}{arg\,max} 

 \providecommand{\norm}[1]{\lVert#1\rVert}
\usepackage{array}

\usepackage{mdwmath}
\usepackage{amsfonts}
\usepackage{mdwtab}
\usepackage{eqparbox}
\usepackage{relsize}
\usepackage{mdframed}
\mdfdefinestyle{review}{linecolor=white,backgroundcolor=yellow!40}
\mdfsetup{skipabove=0pt,skipbelow=0pt}
\ifCLASSOPTIONcompsoc
\usepackage[caption=false,font=normalsize,labelfon
t=sf,textfont=sf]{subfig}
\else
\usepackage[caption=false,font=footnotesize]{subfig}
\fi
\usepackage{url}
\usepackage{CJK}
\usepackage{gensymb}
\newcommand{\bi}{\begin{itemize}}
	\newcommand{\ei}{\end{itemize}}
\newcommand{\bfr}{\begin{frame}}
	\newcommand{\efr}{\end{frame}}

\usepackage{enumitem}
\setlist[itemize]{noitemsep, topsep=0pt}
\makeatletter
\newcommand\subparagraph{%
	\@startsection{subparagraph}{5}
	{\parindent}
	{3.25ex \@plus 1ex \@minus .2ex}
	{-1em}
	{\normalfont\normalsize\bfseries}}
\makeatother
\usepackage{titlesec}
\let\subparagraph\relax 
\titlespacing*{\section}{0pt}{10pt}{7pt}
\titlespacing{\subsection}{0pt}{5pt}{5pt}
\titlespacing{\subsubsection}{0pt}{3pt}{1pt}
\colorlet{bgcolor}{yellow}
\usepackage{soul}

\allowdisplaybreaks[3]

\begin{document}
	\title{ A Stochastic Model for Short-Term Probabilistic Forecast of Solar Photo-Voltaic Power}
	\author{\IEEEauthorblockN{Raksha Ramakrishna, \textit{Student Member, IEEE}, Anna Scaglione, \textit{Fellow, IEEE}, Vijay Vittal, \textit{Fellow, IEEE}}
		\thanks{The authors are with the School of Electrical, Energy and Computer Engineering (ECEE), Arizona State University, Tempe, AZ 85281, USA. This work was funded in part by the Advanced Research Projects Agency- Energy (ARPA-E), U.S. Department of Energy, under Award Number DE-AR0000696 and by National Science Foundation under grant number CPS-1549923 . The views and opinions of authors expressed herein do not necessarily state or reflect those of the United States Government or any agency thereof.}	\vspace{-0.7cm}~}			
	\maketitle	
	\begin{abstract}
			In this paper, a stochastic model with regime switching is developed for solar photo-voltaic (PV) power in order to provide short-term probabilistic forecasts. 
		The proposed  model for solar PV power is physics inspired and explicitly incorporates the stochasticity due to clouds using different parameters addressing the attenuation in power.
		Based on the statistical behavior of parameters, a simple regime-switching process between the three classes of \textit{sunny}, \textit{overcast} and \textit{partly cloudy} is proposed.
		Then, probabilistic forecasts of solar PV power are obtained  by identifying the present regime using PV power measurements and assuming persistence in this regime.
		To illustrate the technique developed, a  set of solar PV power data from a single rooftop installation in California is analyzed and the effectiveness of the model in fitting the data and in providing short-term point and probabilistic forecasts is verified. 
		The proposed forecast method outperforms a variety of reference models that produce point and probabilistic forecasts and therefore portrays the merits of employing the proposed approach.
	\end{abstract}
		
	\section{Introduction}
	Solar power generation, both from PV farms and roof-top solar panel installations is on the rise leading to their increasing penetration into traditional energy markets. Hence, consideration of  solar PV power resource while analyzing electric grid operations is gaining great significance. 
	Accurate models that can not only provide solar power forecasts, but also capture the uncertainty in the random process are necessary to address decision problems such as stochastic optimal power flow (SOPF) \cite{Bazrafshan2016}, probabilistic power flow studies \cite{Probabilistic_power_flow}, designing microgrids \cite{Stadler2016}, solar power shaping \cite{solar_power_shaping} and reserve planning. 
	\par A vast array of literature exists in the area of short-term point forecasts for solar power.  A majority of the approaches taken can be broadly classified as being physical, statistical or a hybrid of the two methods (see e.g. \cite{Lit_review_solar_forecast} for a review). 
	Physical methods employ astronomical relationships \cite{Hottel1976}, meteorological conditions and numerical weather predictions (NWPs) for an improved forecast \cite{Long_term_pattern_30_years,Lorenz2009}. Such studies are based on modeling the clear sky radiation using earth-sun geometry, panel tilt and orientation, temperature and wind speed \cite{Coimbra2016,Hoff2010}. Some also use irradiance information available from databases to determine the value of power for a geographical location considered.
	 Other papers use static images of clouds in the sky recorded by a total sky imager (TSI) \cite{sky_image} or utilize a network of sensors recording cloud motion \cite{cloud_motion_vector} to predict solar power. These models rely on a deterministic mapping  given additional information to produce an estimate of the power generated by the panel. 
  \par	The most prevalent statistical methods for solar power forecasting include  time-series modeling such as using  autoregressive (AR) models \cite{Short_term_forecast,SDEforecast,CARDS_prediction}. One of the advantages of these methods is that they are power data-driven and do not depend on having additional information like the previous literature cited, and are adaptive. However, these methods are designed to model stationary normal processes and assume that some transformation such as dividing by the clear sky power time series makes the series stationary. Such assumptions may not be enough to fully capture the non-stationarity of solar power production. Alternatively, there are other approaches such as autoregressive integrated moving average (ARIMA) and autoregressive with exogenous input (ARX) \cite{Lit_review_solar_forecast} based non-stationary methods for solar power prediction.
	 \par In the same class of statistical methods there exist other works that capture variability in solar PV power \cite{Lave2013,Vittal_letters,dic_learning_solar_power} and black box models like using artificial neural networks (ANN) \cite{fuzzy_model,ANN_review,ANN_with_multiple_days} and support vector machines (SVM) \cite{Jang2016} based pattern matching techniques to predict solar power when class labels are known. Additionally, there are also methods  based on the Markovianity assumption of solar power such as \cite{Callaway2016,MC_forecast,markov_switch_irradiance}  in order to forecast solar PV power. 
	\par Contrary to  solar forecasting methods that provide point forecasts, there exist only a few works in the field of probabilistic forecasting of solar PV power. In \cite{chronological_PV_forecast}, a non-parametric kernel density estimation method is used to fit a probability distribution to PV power. In 
		\cite{probabilistic_forecast_high_order_markov}, a higher order Markov chain is used to characterize solar PV power and operating points based on temperature are also used to classify different PV systems and then Gaussian mixture models (GMM) are used for probabilistic forecasts. 
		One can also consider all of the AR based time series methods since they can essentially be used to obtain probabilistic forecasts. 	
	\par In the proposed methodology the advantages offered by both physical and statistical approaches are exploited. The  proposed model provides a statistical description of stochasticity of the electric power signal that is inspired by the physical behavior of solar PV power, while being completely adaptive. 
	The advantage of going by the modeling approach motivated by the physics of the problem is that it helps in understanding the underlying phenomenon and provides an easier interpretation of the results obtained. Also, solar PV power is non-stationary by nature and this needs to be captured by the modeling technique. 
	\par The model at a macro-level defines a regime switching process \cite{regime_switch_econ} which says that solar irradiance emanates from one of the three classes: \textit{sunny, partly cloudy, overcast}. The stochastic models for \textit{sunny} and \textit{overcast} are simple Gaussian distributions whereas for the \textit{partly cloudy} regime, a hidden Markov model (HMM) is proposed. 
	\par Such an approach simplifies the understanding of temporal variations in solar PV power by examining each regime separately and associating a physical meaning to the hidden states. It is also important to note that no assumption of stationarity is made while describing the regime switching process and no attempt is made to estimate this time-varying transition probability. This is usually not the case in most of the other works. In this manner, the proposed method uniquely captures the non-stationarity in solar power which is not just due to its diurnal structure. 
	\par Prior work in \cite{Asilomar2016} by the authors briefly described in Section \ref{sec:Model} involved the development of a parametric model that was proven to efficiently capture the effect of clouds on solar PV power while providing  a compact representation. In this paper, the prior modeling technique  is utilized and extended to fit a switching process to solar PV power, using which a solar power prediction algorithm to provide short-term probabilistic forecasts is designed. The resulting low order model ensures reduced computational complexity for the proposed algorithm. 
	\par  The key contributions of this paper are:
	\begin{itemize}
		\item The proposition of a regime-switching process for solar PV power that consists of periods that can be classified as \textit{sunny}, \textit{overcast} and \textit{partly cloudy} and development of stochastic models for the three regimes. This is detailed in section \ref{sec:stochastic_models_for_classification}.
		\item A hidden Markov model (HMM)  for the \textit{partly cloudy} regime whose latent states are the support of sparse parameters pertaining to attenuation of power. This is described in subsection \ref{subsec:partly_cloudy_model}.
		\item A change detection algorithm to identify the present regime using  solar power data. No other auxiliary information such as temperature or wind speed is used.
		\item The design and analysis of a computationally efficient online algorithm for short term solar power prediction by employing the switching process and the relevant stochastic models for \textit{sunny}, \textit{overcast} and \textit{partly cloudy} as outlined in section \ref{sec:solar_power_prediction}.
	\end{itemize} 
	\par The prediction results using the proposed method, as seen in subsection \ref{subsec:solar_power_pred_results}, indicate the validity of the approach. In fact, the proposed prediction also outperforms multiple reference models including  \textit{ smart persistence} \cite{smart_persist}, \textit{diurnal persistence}, ANN based prediction method, AR model for the stochastic component of solar PV power and multiple AR models, one each for \textit{sunny}, \textit{partly cloudy} and \textit{overcast} with regime switching (c.f subsection \ref{subsubsec:comaprison_pred}).  
	\par Also, one could complement the proposed method by using weather prediction and cloud imagery as additional information in order to improve the performance of the proposed forecast method. Since stochastic models are available, the method  provides probabilistic forecasts which are very useful while making decisions under uncertainty.
   Section \ref{sec:Conclusions} includes the conclusions and future work.
	\section{Discrete time Model}\label{sec:Model}
\begin{figure}	
		\centering
		\includegraphics[height=0.2\textheight]{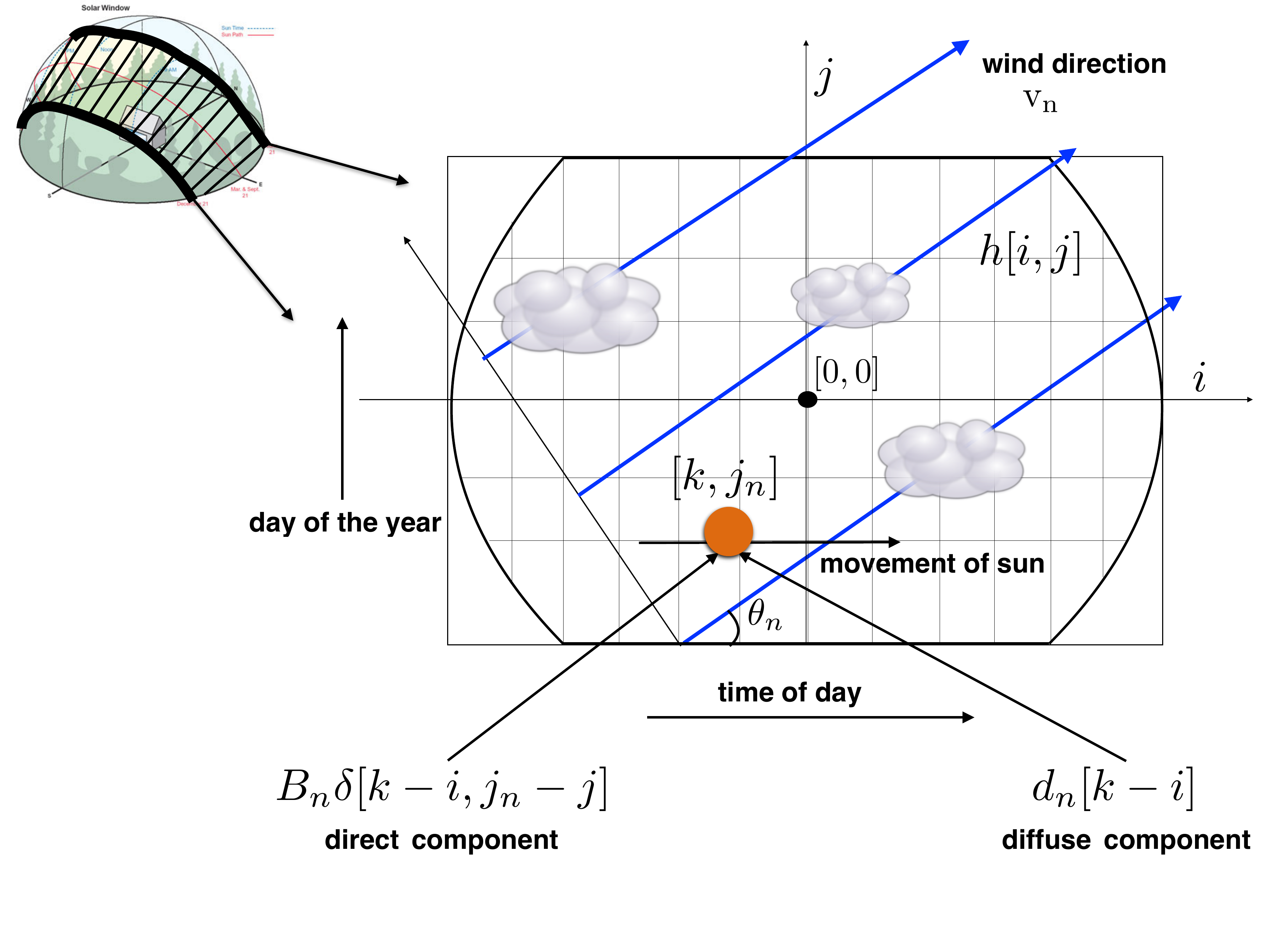}
		\caption{ Figure representing the  sun's path across the sky over the days on a plane. The orange dot marks the position of the sun at time $k$ on day $n$. The wind trajectory is described by blue lines.}
		\label{fig:sun_path}
	\end{figure}

	The following discrete time model for solar PV power output was derived in detail in the authors' prior work in \cite{Asilomar2016}. It hypothesizes that the panel sums solar irradiation from the sky by weighting each contribution with a bi-dimensional gain function that handles the scaling factors to obtain total electrical power. The solar irradiation is attenuated by clouds modeled as a random mask that subtracts a percentage  of the light coming from the patch of sky it covers at a certain time. The motion of the clouds over the panel can be approximated to be moving at a constant speed in a certain direction throughout the day. This assumption is reasonable considering the size of the panel relative to that of the displacement of the clouds. 
	It is known that the solar irradiation has two major components \cite{Masters}, \textit{direct beam component} and \textit{diffuse beam component}. Each of these components is attenuated by the cloud coverage in different ways.
	 Fig.\ref{fig:sun_path} summarizes the idea behind the discrete time model. 
	Let the received solar power be $w_n[k]$ on day $n$ and $k \in (-N,N)$, the discrete time model is given by
	\begin{align}
	w_{n}[k] &= s_{n}[k] -(p_{n}^{b}[k] + p_{n}^{d}[k]) + p_{n}^{e}[k]+ {\eta}_{n}[k] \label{init_model}
	\end{align}	
	where  $s_n[k]$  is the solar power if the $n$th day is sunny, $p_{n}^{b}[k]$ and  $p_{n}^{d}[k]$  are the components pertaining to direct and diffused beam component attenuation by the clouds respectively, $p_{n}^{e}[k]$ is attributed to edge of the cloud effect and ${\eta}_{n}[k]$ is Gaussian measurement noise.
\par In the next two subsections  a parametric model for the solar power output without and with cloud attenuation is provided.  
The goal is to construct stochastic models for each of the three regimes and use them for probabilistic forecasting. Instead of directly formulating these stochastic models, the deterministic model with parameters is first constructed.  Then, the stochasticity in parameters is characterized and further leveraged in section \ref{sec:stochastic_models_for_classification} to define stochastic models for solar power output. 
	\subsubsection{ Sunny days parametrization}\label{subsec:Sunny_days}
\begin{figure}[t]
		\includegraphics[width=\columnwidth,height=0.2\textheight]{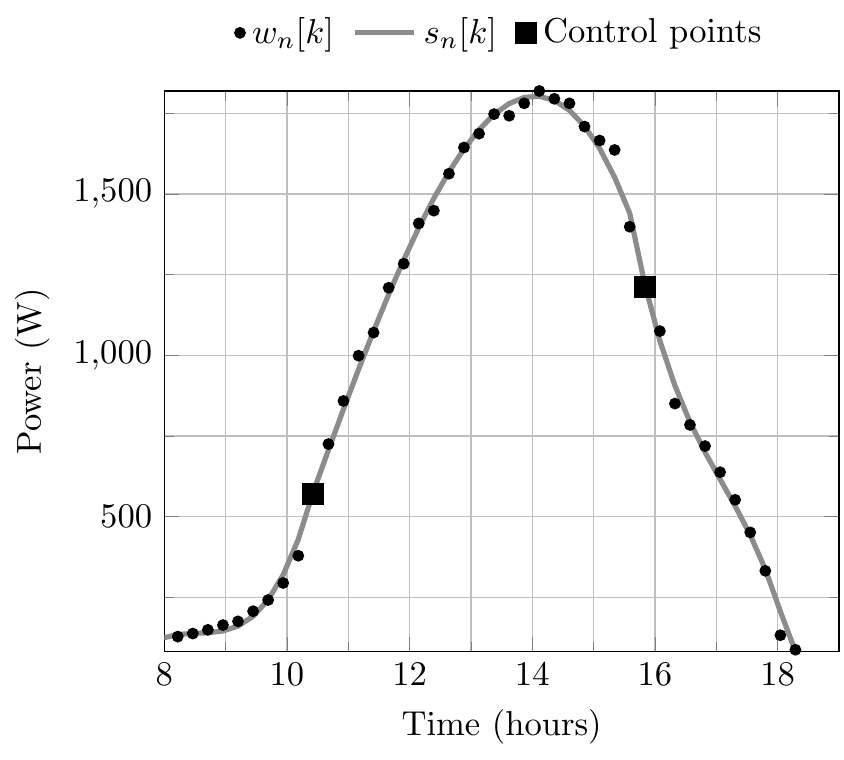}
		\caption{Plot showing $s_{n}[k]$ and $w_{n}[k]$ for a sunny day on October 22, 2009 }
		\label{sunny_fit}
	\end{figure}

	Physics based models give explicit expressions for $s_n[k]$ accounting for the geographical location, orientation and tilt of the panel and time \cite{Masters}. 
	Since each location can have possible variations with shading and a variety of panel orientations, these expressions are not employed. Instead,  each cloudless day is modeled using a simple basis expansion model, whose expansion coefficients are periodically updated to reflect seasonal variations. 
	Let $\mathcal{S}$ denote the set of sunny days. For the $n$th day $n \in \mathcal{S}$ the solar PV power samples are modeled as:
	\begin{align}
	w_n[k]\equiv s_n[k]&=\sum\limits_{q=0}^Q \mathrm{s}_{nq}b_q(k)
	\end{align}
	where the choice of basis is three sets of non-overlapping cubic splines that cover three daylight periods delimited by two control points $k_{n1},k_{n2}$ . The control points are time instants at which there is a discontinuity in the first and second derivative of the signal. For sunny days this is identified numerically from the data. 
	This is shown in Fig.\ref{sunny_fit}.
To constrain the cubic splines covering different periods to have the same values at control points ( $C^{0}$ continuity),  the basis is constructed using $Q=9$ i.e. $10$ functions that are derived from Bernstein polynomials \cite{HermitePoly}, $B_{j,\nu}(t)$ of degree $\nu=3$ as 
	\begin{align}
	b_{q = \nu i + j}(k)&= B_{j,\nu}(t_{i}), ~ i=0,1,2      
	\end{align}	where
	\begin{align}
	B_{j,\nu}(t) &= \binom{\nu}{j} t^{j} (1-t)^{\nu-j} \mbox{rect}(t), ~ j=0,1,2,3
	\end{align}, $\mbox{rect}(t)$ denotes the rectangular function between $[0,1)$ and 
	\begin{align}
	t_{0} &=(k+N)/(k_{n1}+N),& -N \leq k \leq k_{n1}\\
	t_{1}&=(k-k_{n1})/(k_{2n}-k_{1n}),& k_{n1}\leq k \leq k_{n2}\\
	t_{2}&=(k-k_{n2})/(N-k_{n2}),& k_{n2}\leq k \leq N
	\end{align}
	Thus, for a sunny day at most $10$ parameters plus $2$ control points are needed. The approximated $s_{n}[k]$ for one such sunny day is shown in Fig. \ref{sunny_fit}. It highlights the very specific pattern obtained in October due to shading.
	\subsubsection{ Cloudy days parametrization}\label{subsec:Cloudy_days}
	By referring to the authors' previous work \cite{Asilomar2016} where  the expressions for $p_{n}^{b}[k]$ and  $p_{n}^{d}[k]$ were derived as,
	\begin{align}
	p_{n}^{b}[k]  &\approx a_{n}^{b}[k] s_{n}[k],~~  a_{n}^{b}[k] = \sum\limits_{\ell \in \mathcal{B}} a_{\ell}\delta[k-r_{\ell}] \label{eq:direct_beam_attn}	\\
		p_{n}^{d}[k] &\approx \sum\limits_{q} \tilde{h}[q]z_{n}[k-q] \label{eq:diffuse_beam_attn} 
	\end{align}
	and where $a_{n}^{b}[k] $ is the stochastic time series capturing the direct beam sudden power attenuations caused by clouds whose trajectories intersect with that of the sun. The diffuse beam attenuation, instead, is modeled as the convolution of a one-dimensional filter $\tilde{h}[k] $ with a stochastic input $z_{n}[k]$ that represents the cloud attenuation. Furthermore, to explain the increase of power even beyond the expected sunny day power $s_{n}[k]$,  along the lines of direct beam attenuation the following term is introduced to be
present only when $w_{n}[k]>s_{n}[k]$:	\begin{align}
	p_{n}^{e}[k]  \approx a_{n}^{e}[k] s_{n}[k],~~  a_{n}^{e}[k] = \sum\limits_{\ell \in \mathcal{E}} a_{\ell}\delta[k-k_{\ell}]
	\end{align} 
This term captures the so called \textit{edge of the cloud effect} that has been reported in literature \cite{kankiewicz2010observed,edge_of_clouds_curtright}. The edges of some clouds, $\ell \in \mathcal{E}$ act like a magnifying lens when their paths intersect with that of the sun thereby boosting the power. 
	It is important to note that the edge of cloud effect cannot occur simultaneously with cloud related attenuation and in general this term will be far sparser. 
	This observation directly ties to the formulation of the regression problem  presented next. 		
	\subsubsection{Regression problem}\label{sec:complete-model}
	From the cloudy days parametrization, the complete model for power on day $n$ can be written as:
	\begin{equation}
	w_{n}[k] = \begin{cases}
	s_{n}[k](1- a_{n}^{b}[k]) - \sum\limits_{q} \tilde{h}[q]z_{n}[k-q], w_{n}[k]  \leq s_{n}[k] \\
	s_{n}[k] + a_{n}^{e}[k] s_{n}[k], w_{n}[k]  > s_{n}[k]
	\end{cases}\label{eq:model_scalar}
	\end{equation}	
	Equation \eqref{eq:model_scalar} distinguishes between attenuation, $w_{n}[k]  \leq s_{n}[k]$ and the edge of cloud effect, $w_{n}[k]  > s_{n}[k]$.  One can write the convolution term in matrix-vector form with extended end conditions \cite{oppenheim} as $\mathcal{T}(\tilde{\mathbf{h}})\mathbf{z}_{n}$ where $\mathbf{z}_{n}(i) = z_{n}(i-M+1), \mathbf{z}_{n} \in \mathbb{R}^{(2N+M-1) \times 1}_{+}$, $\tilde{\mathbf{h}}(i) = \tilde{h}(i), \tilde{\mathbf{h}} \in \mathbb{R}^{M \times 1}_{+}$ and $\mathcal{T}(\tilde{\mathbf{h}}) \in \mathbb{R}^{2N \times (2N+M-1)}_{+}$  is the Toeplitz matrix with first column $\left[\begin{matrix}
\tilde{h}[M-1], & \mathbf{0}^{1 \times 2N-1}
\end{matrix}\right]^{T}$ and first row $\left[\begin{matrix}
\tilde{h}[M-1],\ldots,\tilde{h}[0],& \mathbf{0}^{1 \times 2N-1}
\end{matrix}\right]$. 
\par Also, the direct beam attenuation term and the edge of cloud effect can be written as $\mathbf{S}_{n}^{b}\mathbf{a}_{n}^{b}$ and $\mathbf{S}_{n}^{b}\mathbf{a}_{n}^{e}$ respectively where $\mathbf{s}_{n}(i) = s_{n}(i) ,\mathbf{a}_{n}^{e}(i) = a_{n}^{e}(i), \mathbf{a}_{n}^{b}(i) = a_{n}^{b}(i)$ are $2N$ dimensional positive real vectors, $\mathbf{S}_{n}^{b} = \textrm{diag}(\mathbf{s}_{n}) \in \mathbb{R}^{2N \times 2N}$. 
	Then, writing \eqref{eq:model_scalar} in vector form for $\mathbf{w}_{n}(i) = w_{n}(i)$,
	\begin{align}
	\mathbf{w}_{n} = \mathbf{s}_{n}-\mathbf{U} (\mathbf{S}_{n}^{b}{\mathbf{a}_{n}^{b}} + {\mathcal{T}(\tilde{\mathbf{h}})\mathbf{z}_{n}}) + \tilde{\mathbf{U}}\mathbf{S}_{n}^{b}\mathbf{a}_{n}^{e}
	\end{align}
	where $U(.)$ is the Heaviside step function operating element-wise,  $\mathbf{U} =  \textrm{diag}( U(\mathbf{s}_n-\mathbf{w}_{n}))$,$ \tilde{\mathbf{U}} =  \mathbb{I}-\mathbf{U}$ and  $\mathbb{I}$ is the identity matrix of size $2N$. 
	Here the estimation of the cloud coverage parameters is seen as a blind deconvolution problem that falls in the class of sparse dictionary learning problems \cite{Dic_learning_lit_review,Shift_inv_dic}, usually solved by alternating between the estimation of the vectors $\mathbf{z}_n,  \mathbf{a}_{n}^{b}, \mathbf{a}_{n}^{e}$ by sparse coding\cite{compressive_sensing_lit_review}  and the  estimation of filter $\tilde{\mathbf{h}}$ over multiple iterations.
	More specifically, as in a typical sparse coding problem formulation, estimates can be obtained by solving: 
	\begin{align} \label{sparse_coding_edge_of_cloud}
	\begin{aligned}
	\!\!\!\!& \min_{\tilde{\mathbf{h}} , \mathbf{z}_{n},\mathbf{a}_{n}^{b},\mathbf{a}_{n}^{e}}
	& &\sum\limits_{n}\norm{\mathbf{U}\left(\mathbf{s}_n-\mathbf{w}_{n}  - \mathbf{S}_{n}^{b}\mathbf{a}_{n}^{b} - {\mathcal{T}}(\tilde{\mathbf{h}})\mathbf{z}_n\right) \\[-2pt]
		&&&+ \tilde{\mathbf{U}}\left(\mathbf{s}_n -\mathbf{w}_{n} + \mathbf{S}_{n}^{b}\mathbf{a}_{n}^{e} \right)}_{2}^{2}\\[-1pt]
	&&&+\sum\limits_{n} \lambda_1 (\mathbf{1}^{T}\mathbf{a}_{n}^{e})  + \lambda_2 (\mathbf{1}^{T}\mathbf{a}_{n}^{b}) + \lambda_3 (\mathbf{1}^{T}\mathbf{z}_{n})\\[-2pt]
	& {\text{subject to}}
	& &  \mathbf{a}_{n}^{b}\geq 0,  ~~,\mathbf{a}_{n}^{e}\geq 0,~~\mathbf{z}_{n} \geq 0 ~~ \forall n, ~~\tilde{\mathbf{h}}\geq 0 \\
	&&&\tilde{\mathbf{U}} \left(\mathbf{S}_{n}^{b}\mathbf{a}_{n}^{b} + {\mathcal{T}}(\tilde{\mathbf{h}})\mathbf{z}_n\right)=\mathbf{0},~~\mathbf{U}\mathbf{S}_{n}^{b}\mathbf{a}_{n}^{e}=\mathbf{0}
	\end{aligned}
	\end{align}
	The algorithm is initialized with the filter being a scaled Hamming window of length $M$, $\tilde{h}[q] = g \times(0.54-0.46\cos\left(2 \pi q/(M-1)\right))$. The alternating algorithm is guaranteed to find only a locally optimal solution and it depends on the initialization. To address the scale ambiguity inherent in blind deconvolution problems, the scale $g$ is chosen such that $\tilde{h}[q]$ and $s_{n}[k]$ have similar amplitudes. Regularization constants are chosen such that $\lambda_{1} \ge \lambda_{2} \gg \lambda_{3}$ in order to force $\mathbf{a}_{n}^{b}$ and $\mathbf{a}_{n}^{e}$ to be more sparse than $\mathbf{z}_{n}$. This rationale is justified since only a subset of total number of clouds have the possibility of directly occluding the sun.	
It was shown in \cite{Asilomar2016} that the proposed model led to an excellent fit with the data. Even though the regression problem is solved in a completely deterministic fashion, such a model allows the separation of the components and study  a plausible stochastic model for them. This is explained in detail in the next section \ref{sec:stochastic_models_for_classification}.
	\section{Stochastic models for classification and forecast of solar power data} \label{sec:stochastic_models_for_classification}
\begin{figure}
			\centering
			\includegraphics[height=.15\textheight]{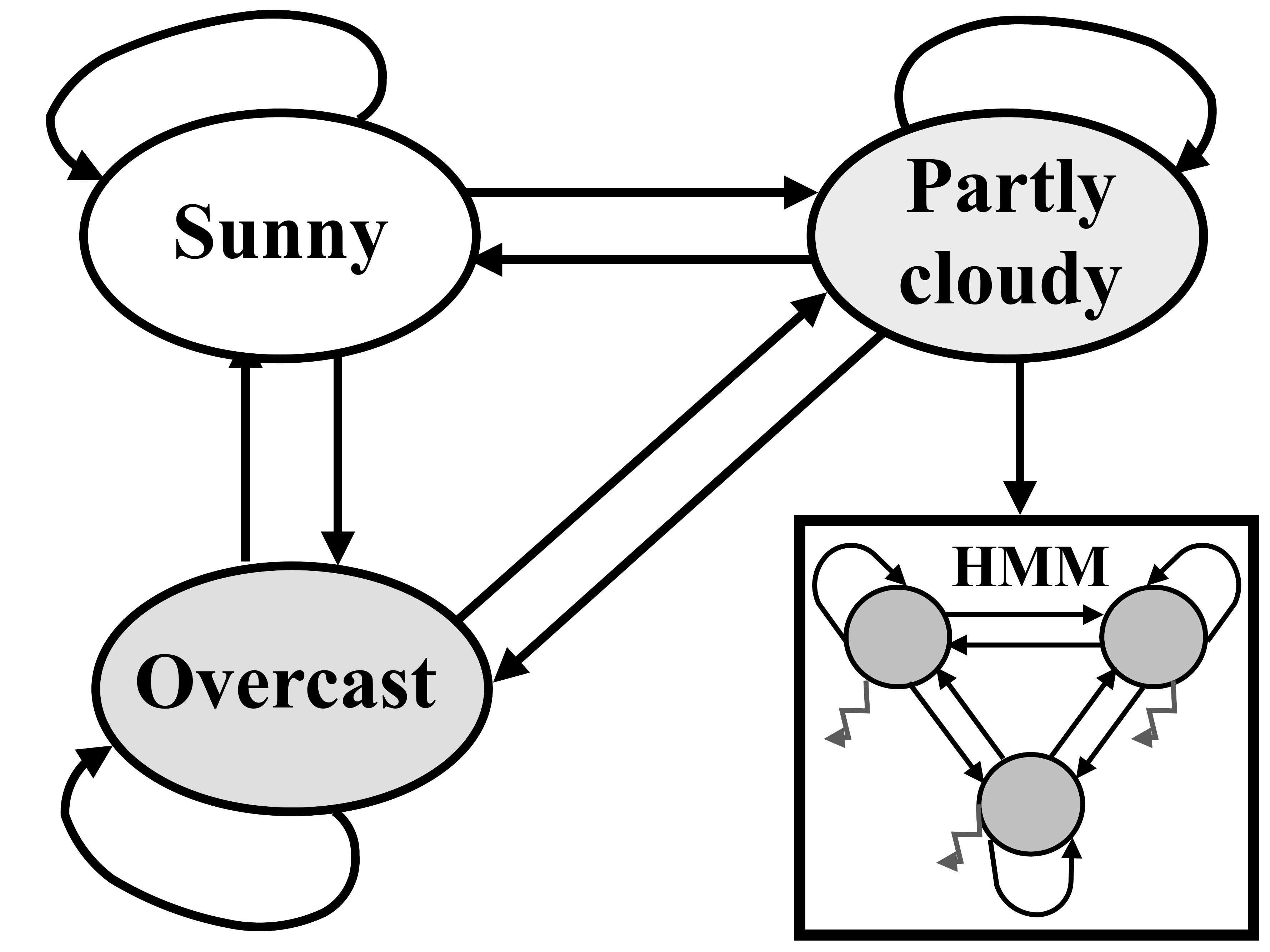}
			\caption{Block diagram highlighting the proposed switching process between stochastic models}
			\label{fig:stochastic_model_block_diagram}
		\end{figure}

	In spite of the fact that the switching nature is not intrinsically part of the model discussed above, as reported by the authors in \cite{Asilomar2016},  the results of the deterministic fit after solving \eqref{sparse_coding_edge_of_cloud} highlighted the switching nature of the solar irradiation phenomenon. Solar PV power produced in a period of time can be broadly classified as coming from \textit{sunny, overcast} or \textit{partly cloudy}	models. The model switches between the three classes as shown in Fig. \ref{fig:stochastic_model_block_diagram} due to weather changes.  In this section, a stochastic model for each of the three classes is proposed. The first application of this model is for {\it change detection}, i.e. to identify the switch between classes to provide a forecast  by  assuming that the model persists. The second application is for probabilistic short-term forecast.   	
	\subsubsection{Stochastic model for sunny period} \label{subsec:sunny_period}
	For sunny periods, it is hypothesized that the solar power  is the deterministic solar power pattern i.e., 
	\begin{align}
	{w}_{n}[k] = {s}_{n}[k] + {\eta}_{n}[k]
	\end{align}
	The modeling error is given by ${\eta}_{n}[k] \sim \mathcal{N} (0, \sigma_{s}^2 ) ~ \forall k$ and $ n \in \mathcal{S}$. The variance $\sigma^{2}_{s}$ is estimated using the error values after fitting the sunny day pattern from section.\ref{sec:Model} to the sunny days, $n \in \mathcal{S}$
	\subsubsection{Stochastic model for overcast period}\label{subsec:overcast_period}
	During overcast periods, the attenuation of solar power is mostly from the diffuse beam component \cite{Masters} which is why ${z}_{n}[k]$ accounts for the relevant attenuation. Also, there is an average component in the overcast days for ${z}_{n}[k]$ that mimics a scaled version of sunny day pattern ${s}_n[k]$. Since this attenuation is smooth, the model for overcast period is:
	\begin{align}
	{w}_{n}[k] \approx \alpha_{n}{s}_{n}[k] + {\eta}_{n}[k] ,  \label{eq:overcast_approx} \\
	\eta_{n}[k] \sim \mathcal{N}(0,\sigma_{oc}^{2})
	\end{align} 
	where $\alpha_{n}$ can be thought of as the attenuation of sunny day power. The parameter $\alpha_{n}$  is  analogous to \textit{clear sky index} defined as $w_{n}[k]/s_{n}[k]$ that many papers use to model solar PV power \cite{SolarForecastCoimbra2013}. However, all the samples in the overcast period are used to estimate $\alpha_{n}$ unlike the determination of clear sky index. This leads to robustness with respect to noise. 
	\par In order to limit the values of power ${w}_{n}[k] $ to values between $0$ and $s_{n}[k]$, a truncated Gaussian distribution is considered for $w_{n}[k]$,
	\begin{align}
  \tilde{f}_{oc}({w}_{n}[k]) = \frac{(1/\sqrt{2\pi \sigma_{oc}^{2}}) \exp{\left((w_{n}[k] - \alpha s_{n}[k])^{2}/ 2 \sigma_{oc}^{2}\right)}}{\Phi(\frac{s_{n}[k] - \alpha{s}_{n}[k]}{\sigma_{oc}}) - \Phi(\frac{-\alpha {s}_{n}[k]}{\sigma_{oc}})} \label{oc_PDF}
	\end{align}
	where $\Phi(.)$ denotes the CDF of a standard normal distribution.
	\subsubsection{Stochastic model for partly cloudy period}\label{subsec:partly_cloudy_model}
	The model for the partly cloudy period is slightly more involved, due to the presence of all the three parameters.  However, a hidden Markov model (HMM) is able to capture the underlying on-off process that characterizes the sparse parameters in periods with fast moving clouds that cause sharp fluctuations in solar PV power.
\par The observed solar PV power data ${w}_{n}[k] $  is modeled as coming from underlying hidden states that are Markovian in nature. Let the state/latent variable in this model, $\mathbf{q}_{k}$  be the support of the unknown sparse parameters, $({z}_{n}[k], a_{n}^{b}[k], a_{n}^{e}[k])$. Their relationship is governed by the following equations:
		\begin{align}
		{w}_{n}[k] &= {s}_{n}[k] - \mathbf{P}~\text{diag}\left(\boldsymbol{\Phi} \mathbf{q}_{k}\right)\mathbf{x}_{k} \label{HMM:obs}\\ 
		\mathbf{q}_{k+1} &= \mathbf{A}^{T}\mathbf{q}_{k} + \boldsymbol{\nu}_{k+1}  \label{HMM:state}
		\end{align}
		\begin{align}
	  \text{where}~	\mathbf{P} &= \left[ \begin{matrix}
		\tilde{h}[M-1] &\ldots&\tilde{h}[0]& {s}_{n}[k] & -{s}_{n}[k] 
		\end{matrix}\right],\\
		\mathbf{x}_{k} &= \left[\begin{matrix} 
		{z}_{n}[k-M+1] & \dots & {z}_{n}[k] & {a}_{n}^{b}[k] & {a}_{n}^{e}[k]
		\end{matrix}\right]^{\textbf{T}}
		\end{align} 
		Let the total number of states be represented by $\mathcal{N}_{s}$. Then, $\mathbf{A} \in \mathbb{R}^{(\mathcal{N}_{s} \times \mathcal{N}_{s})}$ is the state transition matrix where $\mathbf{A}(i,j)$ is the probability of going from state $i$ to state $j$ and $\boldsymbol{\nu}_{k+1}$ is the noise.
		The state vector $\mathbf{q}_{k} \in \mathbb{R}^{({M+2}) \times 1}$ is a binary vector taking values from the set of coordinate vectors $\{\mathbf{e}_{1},\mathbf{e}_{2},\dots,\mathbf{e}_{\mathcal{N}_{s}} \}$ where $\mathbf{e}_{i} \in \mathcal{R}^{\mathcal{N}_{s}}$ has a $1$ at  position $i$ and zero elsewhere. The matrix $\boldsymbol{\Phi} \in \mathbb{R}^{(M+2) \times \mathcal{N}_{s}}$ contains the possible combinations of presence and absence of the coefficients in $\mathbf{x}_{k}$ where each combination corresponds to one state. 
		Certain assumptions are made to decrease the number of states.  Firstly,  $\left[\begin{matrix}{z}_{n}[k-M+1] & \dots & {z}_{n}[k]\end{matrix}\right]$ is restricted to have $\ell < M$ non-zero entries. 
		 Secondly, ${a}_{n}^{e}[k]$ cannot co-exist with the other parameters due to the fact that edge of cloud effect is indicative of the absence of attenuation.
		 Furthermore, as a simplification, it is also assumed that direct beam and diffuse beam attenuations do not occur together which means that the total number of states is 
		 \begin{align}
		 \mathcal{N}_s = \sum\limits_{\tilde{\ell}=0}^{\ell} \left(\begin{matrix}
		 M \\ \tilde{\ell}
		 \end{matrix}\right) + 2
		 \end{align}	
		Notice the absence of noise term in the observation equation \eqref{HMM:obs}. This stems from the fact that  measurement noise  is not included since the `noisy' nature of the solar power data is caused by the fast movement of clouds rather than by erroneous measurements. \\
	     The simplest case of choosing $\ell=1$ and having $\mathcal{N}_{s} = M+3$ states is considered. 
		 All non-zero parameters in $\mathbf{x}_{k}$ are hypothesized to come from independent exponential distributions. While in state $i$  a certain ${w}_{n}[k]$ is observed: 
		\begin{align}
		{w}_{n}[k] = \begin{cases}
		{s}_{n}[k], ~ i=1 \\
		{s}_{n}[k] - \tilde{h}[i-2] {z}_{n}[k-i+2], 
		~i = {2,\dots,M+1},\\
		{s}_{n}[k] - {s}_{n}[k] {a}_{n}^{b}[k] ,~ i=M+2\\
		{s}_{n}[k] + {s}_{n}[k] {a}_{n}^{e}[k],~ i=\mathcal{N}_{s}
		\end{cases} 
		\end{align}		
		The corresponding conditional probability distribution  given the state $i$ is denoted as $\tilde{f}_{i}({w}_{n}[k]) \triangleq f_{\underaccent{\tilde}{{w}}_{n}[k]}({w}_{n}[k] | \mathbf{q}_{k} = \mathbf{e}_{i})$ and is equal to,
		\begin{align}
		\tilde{f}_{i}({w}_{n}[k]) =
		\begin{cases}
		\delta({s}_{n}[k]-{w}_{n}[k]), ~ i=1 \\
		 \frac{C_{i} \lambda_{z}}{\tilde{h}[i-2]} \exp{\left\{- \frac{\lambda_{z}\left({s}_{n}[k]-{w}_{n}[k]\right)}{\tilde{h}[i-2]} \right\}} 
		i = {2,\dots,M+1}\\
		\frac{C_{i} \lambda_{a}^{b}}{{s}_{n}[k]} \exp{\left\{- \frac{\lambda_{a}^{b}}{{s}_{n}[k]} \left({s}_{n}[k]-{w}_{n}[k]\right)\right\}},i=M+2 \\
		\frac{\lambda_{a}^{e}}{{s}_{n}[k]} \exp{\left\{- \frac{\lambda_{a}^{e}}{{s}_{n}[k]} \left({w}_{n}[k]-{s}_{n}[k]\right)\right\}},i=\mathcal{N}_{s} \\
		\end{cases}  \label{cond_PDF}
		\end{align}
		where $C_{i}$ is the normalizing constant for the probability distribution given by
		\begin{align}
		C_{i}^{-1}= \begin{cases}
		1-\exp{\{-\lambda_z {s}_{n}[k] /h[i-2] \}}, ~ i= 2,3,\dots M+1\\
		1-\exp{(-\lambda_{a}^{b})}, ~ i=M+2 
		\end{cases}.
		\end{align} The normalization is done so that  ${w}_{n}[k] \in [0, {s}_{n}[k]]$. 		
		\subsection{Learning the parameters of  HMM for partly cloudy periods}\label{subsubsec:learning_HMM}
\begin{figure}[t]
		\centering
		\includegraphics[height=0.2\textheight]{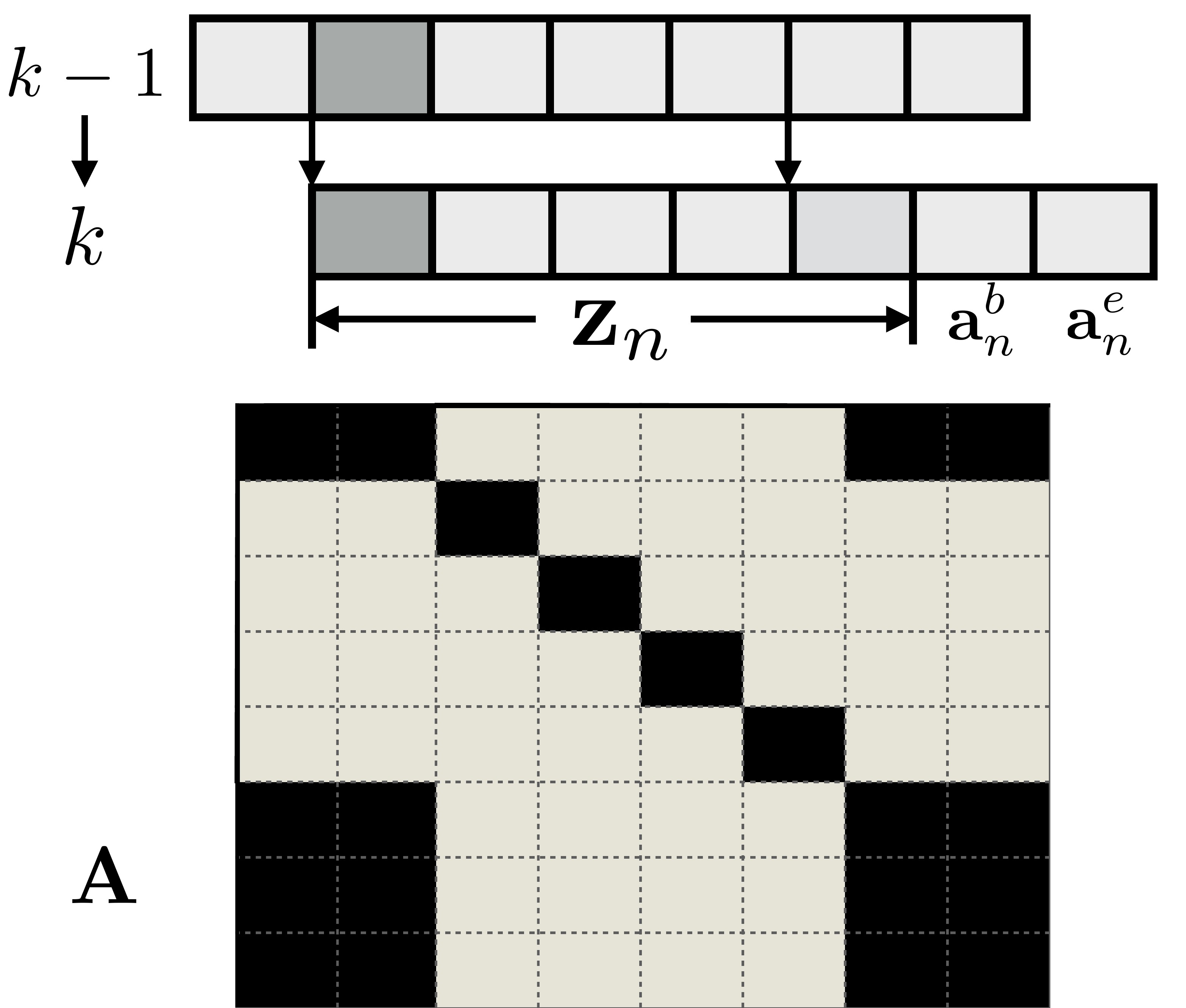}
		\caption{The specific way in which state transition from time instant $k-1$ to $k$ takes places determines the structure of the state transition matrix $\mathbf{A}$}
		\label{fig:state_transition_matrix}
	\end{figure}

		The models for sunny and the overcast periods are such that the only thing that can be predicted is the mean of the process in both cases, but not the noise $\eta_n[k]$ which by construction is assumed to be i.i.d. during the corresponding period. Hence, the problem of learning the stochastic parameters of the model to perform predictions is non-trivial only during partly cloudy periods.   
		To do so, it  is assumed that the values of the parameters $\lambda_z,\lambda_a^{b},\lambda_{a}^{e}$ of conditional probability distributions are known. 	
		It was seen that the algorithm is not very sensitive to the exact values of these parameters as long as they follow ${\lambda}_{z} \leq {\lambda}_{{a}^{b}} \leq \lambda_{a}^{e}$ which is consistent with the results of the regression problem. The probability of starting from a state $i$ denoted by $\pi_{i} = 1/\mathcal{N}_{s}$  is also assumed to be known. 
		In order to learn the the  state transition matrix $\mathbf{A}$, Viterbi training \cite{VT_Jelinek} or segmental k-means \cite{segmental_k_means_Rabiner} approach was adopted. 
		Let $\xi = \{\mathbf{A}(i,j)| i,j \in \{1,\mathcal{N}_{s}\}\}$ be the set of unknown parameters to be estimated. Let $\tilde{N}$ be the  number of samples in a certain block of solar PV power data,  sequence $\mathbf{Q} = \mathbf{q}_{1},\mathbf{q}_{2},\dots,\mathbf{q}_{\tilde{N}}$ and $\mathbf{W}={w}_{n}[1],{w}_{n}[2],\dots,{w}_{n}[\tilde{N}] $ denote the sequence of solar power observations. 
	In the Viterbi training algorithm, instead of maximizing the likelihood over all possible state sequences $\bar{\mathcal{Q}}$,   the likelihood is maximized only over  the most probable state sequence to  find the estimates of parameters in ${\xi} $. The algorithm starts with an  initial estimate for all the unknown parameters $\xi_{0} = \{\mathbf{A}_{0}(i,j)| i,j \in \{1,\mathcal{N}_{s}\}\}$ and performs this maximization iteratively \cite{segmental_k_means_Rabiner},
	\begin{align}
	\hat{\xi}_{m} = \argmax_{\xi} \left( \max_{\mathbf{Q}} f(\mathbf{W},\mathbf{Q}|\hat{\xi}_{m-1})\right)
	\end{align} 
where $m$ is the iteration number and
 \begin{align}
 f(\mathbf{W},\mathbf{Q}|\xi) = p(\mathbf{q}_{1}) \prod_{k = 1}^{\tilde{N}} p({w}_{n}[k] | \mathbf{q}_{k}, \xi) \prod_{k=1}^{\tilde{N}-1}p(\mathbf{q}_{k+1}|\mathbf{q}_{k},\xi) 
 \end{align}
 The inner maximization is performed by using a  dynamic programming algorithm known as Viterbi algorithm \cite{viterbi_algorithm} which is a  recursive method. As a result of this maximization,
 \begin{align}
 \mathbf{\hat{Q}}_{m} &= \argmax_{\mathbf{Q}} f(\mathbf{W},\mathbf{Q}|\hat{\xi}_{m-1}) =\mathbf{\hat{q}}^{m}_{1},\mathbf{\hat{q}}^{m}_{2},\dots,\mathbf{\hat{q}}^{m}_{\tilde{N}}, 
 \end{align} the most likely state sequence at iteration $m$ which best describes the observed data. Later, maximum-likelihood (ML) estimates $\hat{\xi}_{m}$ are estimated,
	\begin{align}
	\hat{\xi}_{m} = \argmax_{\xi} \left(  f(\mathbf{W},\mathbf{\hat{Q}}_{m}|\hat{\xi}_{m-1})\right)
	\end{align} 
	Maximizing $\log{f(\mathbf{W},\mathbf{\hat{Q}}_{m}|\hat{\xi}_{m-1})}$ with respect to $\mathbf{A}_{m}(i,j)$ under the constraint that $\mathbf{A}_{m}$ is stochastic since it is the state transition matrix i.e.,
	$\sum\limits_{j=1}^{\mathcal{N}_{s}} \mathbf{A}_{m}(i,j) = 1 $, gives
			\begin{align}
		\hat{\mathbf{A}}_{m}(i,j) = \frac{N_{ij}}{\sum\limits_{j=1}^{\mathcal{N}_{s}} N_{ij}}
			\end{align} 
   where $N_{ij}$ is the number of times the transition from state $i$ to state $j$ occurs within the state sequence $\mathbf{\hat{Q}}_{m}$. Following from \ref{subsec:partly_cloudy_model} wherein the number of active coefficients at time $k$ in $\mathbf{x}_{k}$ is restricted to $1$, only a limited number of transitions from state $i$ are possible and not to all $\mathcal{N}_{s}$ states. Also, since ${z}_{n}$ is the input to a filter with memory $M$, it means that $M-1$ components need to be retained and shifted while a new one comes in. All of the above reasons give the state transition matrix $\mathbf{A}$ a sparse and specific structure as shown in Fig.\ref{fig:state_transition_matrix} which is forced on $\mathbf{A}_{0}$ during the initialization .
   As a result, only $(M-1) + 4\times 3$ entries of the matrix need to be estimated when $\ell=1$ instead of $(\mathcal{N}_{s})^{2}$.
			\section{Change Detection and Solar power prediction}\label{sec:solar_power_prediction}
		The premise for prediction is the persistence in the weather condition for the time horizon over which a prediction of solar power is provided. Therefore, the proposed prediction algorithm has two steps:
			\begin{itemize}
				\item Classification of the solar power from a given period as coming from one of the three classes of models: \textit{sunny, overcast, partly cloudy}
				\item Assuming that this weather condition persists for the duration of the prediction horizon and provide with a point forecast corresponding to the class  decided in the classification step. 
			\end{itemize}
			Such a scheme captures the inherent switching behavior that solar power exhibits i.e. that of going from one model to another while persisting for a certain duration in each of these. 
			Note that the classification step can be skipped  if prior knowledge in the form of weather prediction is available. 
			The prediction algorithm utilizes a rolling horizon wherein prediction is improved as more data comes in. 
           \subsection{Classification algorithm for solar power}
           The classification algorithm uses the stochastic models for the solar power data as detailed in \ref{sec:stochastic_models_for_classification}. Let $w_{n}[k], ~ k \in (\kappa_{1},\kappa_{2})$ be the solar power samples that have to be classified. It is easy to decide in favor of \textit{sunny} model by computing the error, $\sum({w}_{n}[k] - {s}_{n}[k])^{2}$. If it is less than some power threshold $\tilde{p} = \mu \sigma_{s}$, $\mu>1$, then it is classified as a sunny period. If that is not the case,  the hypotheses \textit{overcast} ($\mathcal{H}_{0}$) or \textit{partly cloudy} ($\mathcal{H}_{1}$) are tested. 
        \begin{align}
        \mathcal{H}_{0}: & {w}_{n}[k] = \alpha_{n}^{\kappa}{s}_{n}[k] +{\eta}_{n}[k], ~ k \in (\kappa_{1},\kappa_{2})  \nonumber\\
         \mathcal{H}_{1}: & {w}_{n}[k] = {s}_{n}[k] - \mathbf{P}~\text{diag}\left(\boldsymbol{\Phi} \mathbf{q}_{k}\right)\mathbf{x}_{k} \nonumber\\ 
         &\mathbf{q}_{k+1} = \mathbf{\hat{A}}^{T}\mathbf{q}_{k} + \boldsymbol{\nu}_{k} ,~~ k \in (\kappa_{1},\kappa_{2}) \nonumber
        \end{align}   
         \begin{align}
        \text{Let}~ \mathbf{W}_{\kappa} &= {w}_{n}[\kappa_{1}],\dots,{w}_{n}[\kappa_{2}],~\text{and}~
         \mathbf{Q}_{\kappa}= \mathbf{q}_{\kappa_{1}},\dots,\mathbf{q}_{\kappa_{2}} \nonumber
         \end{align}.
      It is a composite hypothesis testing problem since $\alpha_{n}^{\kappa}$ is unknown. 
       The maximum likelihood estimate of $\hat{\alpha}_{n}^{\kappa}$ is, 
       \begin{align}
       \hat{\alpha}_{n}^{\kappa} =\frac {\sum\limits_{k=\kappa_{1}}^{\kappa_{2}}{\left({w}_{n}[k] {s}_{n}[k] \right)^{2}}} {\sum\limits_{k=\kappa_{1}}^{\kappa_{2}}{({s}_{n}[k])^{2}}}. \label{eq:z_opt_alpa}
       \end{align}
    Generalized likelihood ratio is not computed. Instead the error,
    \begin{align}
    \sum\limits_{k=\kappa_{1}}^{\kappa_{2}}(\eta_{n}[k])^2 = \sum\limits_{k=\kappa_{1}}^{\kappa_{2}} (w_{n}[k] - \hat{\alpha}_{n}^{\kappa} s_{n}[k])^{2} \label{eq:eta_opt}
    \end{align} is compared with a predefined threshold and also the value of $\alpha_{n}$ is compared with a heuristically set threshold. These rules decide the classification of data as \textit{overcast} model or \textit{partly cloudy}. If the decision is in favor of \textit{partly cloudy}, the  most likely state sequence $\mathbf{Q}_{\kappa}$ that generated the power observations $\mathbf{W}_{\kappa}$ is determined using the Viterbi algorithm with state transition matrix $\mathbf{\hat{A}}$.
%
           \subsection{Prediction for each class of model}
          Based on the classification results on $w_{n}[k], ~ k \in (\kappa_{1},\kappa_{2})$, a solar power forecast, $\hat{w}_{n}[k], k \in (\kappa_{2}+1,\kappa_{2}+\chi)$ is provided. Here, $\chi$ is the length of the prediction horizon.
           \subsubsection{Prediction using sunny model}
           When the detection algorithm chooses the hypothesis that the current solar power data is from a \textit{sunny} model, then:
           \begin{align}
           \hat{w}_{n}[k] = s_{n}[k], ~\forall k \in \{ \kappa_{2}+1, \kappa_{2}+2,\dots,\kappa_{2}+\chi\}
           \end{align}
           Note that the deterministic sequence of the sunny day solar power pattern is known beforehand, and it is updated at a very slow pace on days that are classified as being sunny, to adjust for seasonal variations. 
           \subsubsection{Prediction  using overcast model}
           When the test on $\sum\limits_{k=\kappa_{1}}^{\kappa_{2}}(\eta_{n}[k])^2$ and $\hat{\alpha}_{n}$ decides that hypothesis $\mathcal{H}_{0}$ is true in the duration $k \in (\kappa_{2}\kappa_{1})$, then:           
           \begin{align}
           \hat{w}_{n}[k] = \hat{\alpha}_{n}^{\kappa}s_{n}[k],
            ~\forall k \in \{ \kappa_{2}+1, \kappa_{2}+2,\dots,\kappa_{2}+\chi\} 
           \end{align}
           where $\hat{\alpha}_{n}$ is estimated using \eqref{eq:z_opt_alpa}.
			\subsubsection{Prediction using partly cloudy model}
		Since solar PV power on a partly cloudy day has an underlying Markov Model,  the estimated state transition matrix 
			$\mathbf{\hat{A}}$ is used to determine the most likely future state sequence: $\mathbf{Q}_{pred} \triangleq \mathbf{\hat{q}}_{\kappa_{2}+1}, \dots \mathbf{\hat{q}}_{\kappa_{2}+\chi}$ as
			\begin{align}
			\mathbf{Q}_{pred} &= \max_{\mathbf{Q}} \left(p(\mathbf{q}_{1})  \prod_{k=\kappa_{2}+1}^{\kappa_{2}+\chi-1}p(\mathbf{q}_{k+1}|\mathbf{q}_{k},\xi)\right)
			\end{align}
			by using a modified Viterbi algorithm:
			Define 
			\begin{align}
			\zeta_{k}(i) = \max_{\mathbf{q}_{1},\mathbf{q}_{2},\dots,\mathbf{q}_{k-1}}\left( p(\mathbf{q}_{1},\mathbf{q}_{2},\dots,\mathbf{q}_{k}=\mathbf{e}_{i})\right) \nonumber
			\end{align}
			Then, 
			\begin{align}
			\zeta_{k+1}(j) = \max_{i} \zeta_{k}(i)a_{ij} \nonumber
			\end{align}
			Let $\tilde{j}$ is the last seen state before prediction started i.e. $\mathbf{q}_{\kappa_{2}} =\mathbf{e}_{\tilde{j}} $. The recursion is:
			\begin{align}
		   \zeta_{k}(j) = 	\max_{1 \leq i \leq \mathcal{N}_{s}} \zeta_{k-1}(i)a_{ij},~~
		   \psi_{k}(j)= \argmax_{1 \leq i \leq \mathcal{N}_{s}} \zeta_{k-1}(i) a_{ij} \nonumber
			\end{align}
			with the initialization: 
			\begin{align}
			\zeta_{1}(i) = 1 ~ \forall i = 1,2,\dots,\mathcal{N}_{s},~~
			\psi_{1}(i) = \tilde{j} \nonumber
			\end{align}
			and termination at:
			\begin{align}
			j^{*}_{ \kappa_{2}+\chi}= \argmax_{1 \leq i \leq \mathcal{N}_{s}} \zeta_{\kappa_{2}+\chi}(i) .\nonumber
			\end{align}
			At this point the state sequence backtracking is:
			\begin{align}
			j^{*}_{ k} = \psi_{k+1}(j^{*}_{k+1}), ~ k =  \kappa_{2}+1, \kappa_{2}+2,\dots,\kappa_{2}+\chi-1 \nonumber
			\end{align}
				After $\mathbf{Q}_{pred}$ is determined,  an estimate of vector $\mathbf{\hat{x}}_{k}$,
				\begin{align}
					\mathbf{\hat{x}}_{k} &= \left[\begin{matrix} 
					\hat{z}_{n} & \dots & \hat{z}_{n} & \hat{a}_{n}^{b} & \hat{a}_{n}^{e}
					\end{matrix}\right]^{\textbf{T}} ~ \text{where}\label{eq:x_estimate} 
				\end{align} is created to generate a point prediction. 
				For that purpose, the estimate $\hat{z}_{n}$ for diffuse beam attenuation is obtained from present power measurements which are emissions of the hidden states $i=2,\dots,M+1$ which implies the presence of diffuse beam attenuation,
				\begin{align}
				\hat{z}_{n} &=  \argmin_{z_{n}}  \sum\limits_{k \in \mathcal{L}, i \in \mathcal{B}}\left({w}_{n}[k]-s_{n}[k] +\tilde{h}[i-1] z_{n}\right)^{2}, \nonumber\\
				& {\text{subject to}}~ {z_{n}} \geq 0 \nonumber\\
				\mathcal{L} &= \{k \mid \mathbf{q}_{k} = \mathbf{e}_{i = 2,\dots,M+1}, k \in (\kappa_{1},\kappa_{2})\},\nonumber\\
				\mathcal{B} &= \{i \mid \mathbf{q}_{k}(i+1)=1, k \in (\kappa_{1},\kappa_{2}), 1 <i\leq M+1\} \label{eq:z_n_estimate}
				\end{align}
				This is equivalent to estimating the size and intensity of one single cloud that is responsible for the diffuse beam attenuation in the time frame considered, and is hence retained in the prediction to account for the future attenuation in power.
				\par The estimates of $\hat{a}_{n}^{b}$ and $\hat{a}_{n}^{e}$ are more heuristic however.  This is due to the fact that these parameters are responsible for the sudden and sharp transition in the value of power and it is very difficult to predict them. Therefore, the  values of $\hat{a}_{n}^{b}$ and $\hat{a}_{n}^{e}$ are adjusted in a way so that, $\hat{w}_{n}[k] = \hat{\alpha}_{n}^{\kappa} s_{n}[k]$ when $\mathbf{q}_{k} = \mathbf{e}_{i=M+1,M+2}$. However, whenever $\hat{\alpha}_{n}^{\kappa} <1$ when state $i=M+2$, the parameters $\hat{a}_{n}^{b}$ and $\hat{a}_{n}^{e}$  are replaced with their mean values. 
	       	\begin{align}
				\hat{a}_{n}^{b} = \begin{cases}
				1 - \hat{\alpha}_{n}^{\kappa} ,~ \hat{\alpha}_{n}^{\kappa} <1\\
				1/\lambda_{a}^{b} ,~\text{otherwise}
				\end{cases}  
				\hat{a}_{n}^{e} = \begin{cases}
				\hat{\alpha}_{n}^{\kappa}-1, ~ \hat{\alpha}_{n}^{\kappa} >1 \\
				1/\lambda_{a}^{e}, ~\text{otherwise}
				\end{cases}\label{eq:a_n_estimate}
				\end{align}			
				Then, from \eqref{HMM:obs}, the prediction of solar power is given by,
				\begin{align}
				\hat{w}_{n}[k] = {s}_{n}[k] - \mathbf{P}~\text{diag}\left(\boldsymbol{\Phi} \mathbf{\hat{q}}_{k}\right)\mathbf{\hat{x}}_{k} , ~~k \in ( \kappa_{2}+1,\kappa_{2}+\chi). 
				\end{align}
				\par
\begin{figure}[t]
				\centering
				\includegraphics[width=\columnwidth,height=.2\textheight]{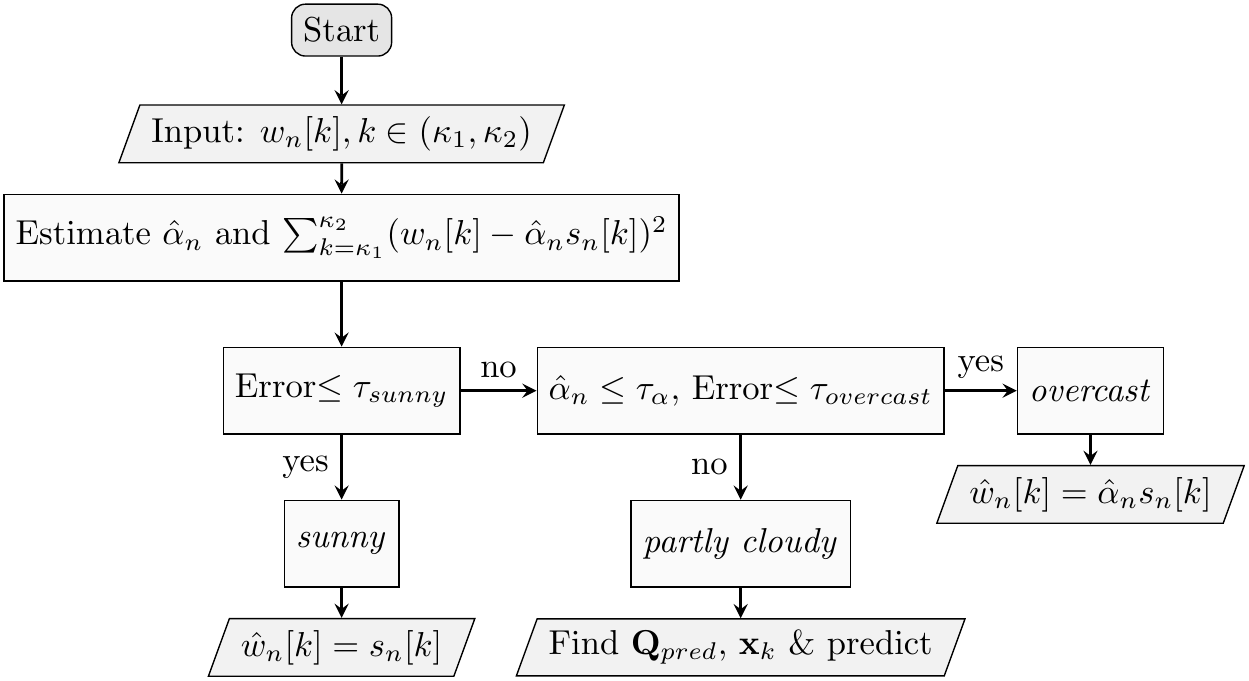}
				\caption{Flowchart of the solar power prediction algorithm}
				\label{fig:solar_power_pred}
			\end{figure}

 The prediction algorithm is summarized in Fig.\ref{fig:solar_power_pred}. 
				\subsubsection{Computational complexity}
			      The computational complexity of the entire prediction algorithm can be calculated as follows:
				To determine the current class/regime, the complexity is that of solving a least-squares problem whose computational complexity is of the order of $\mathcal{O}(\kappa_{2}-\kappa_{1} +1 )$. Then, within the  \textit{partly cloudy} regime, the complexity is mainly due to the Viterbi algorithm and is of the order of $\mathcal{O}(\mathcal{N}^{2}_{s}(\kappa_{2}-\kappa_{1}))$ \cite{viterbi_algorithm}. In order to make a prediction in the \textit{partly cloudy} regime, an additional number of computations is required. The order depends on the prediction horizon, $\chi$. 
				\par Therefore, the computational complexity of the algorithm is of the order of 
				\begin{align}
				\mathcal{O}\left(\kappa_{2}-\kappa_{1} +1 + \mathcal{N}^{2}_{s}(\kappa_{2}-\kappa_{1} + \chi) \right)
				\end{align} using $\chi_{window} = \kappa_{2}-\kappa_{1}$ samples for a prediction horizon of $\chi$. As one can notice, the order is linear in the length of the prediction horizon which is desirable to keep the algorithm computationally efficient. 
	\section{Numerical Results}\label{sec:Results}
	\subsection{Description of the dataset}\label{subsec:fitting_results}
	The dataset used for this work was from a rooftop panel installation in Antioch, California and was provided by SolarCity. This dataset was also used in authors' prior work in \cite{Asilomar2016}. The format of this solar power data consisted of current (in A), voltage measurements (in V) and timestamps (in Hours) at the inverter approximately every $15$ minutes recorded for a duration of two years. Each panel had a rating of $170$ W and there were a total of $22$ panels. Therefore, the nameplate rating of all panels combined was $170 \times 22 = 3740$ W. 
	\begin{figure}[t]
		\centering
	\includegraphics[width=\columnwidth,height=.13\textheight]{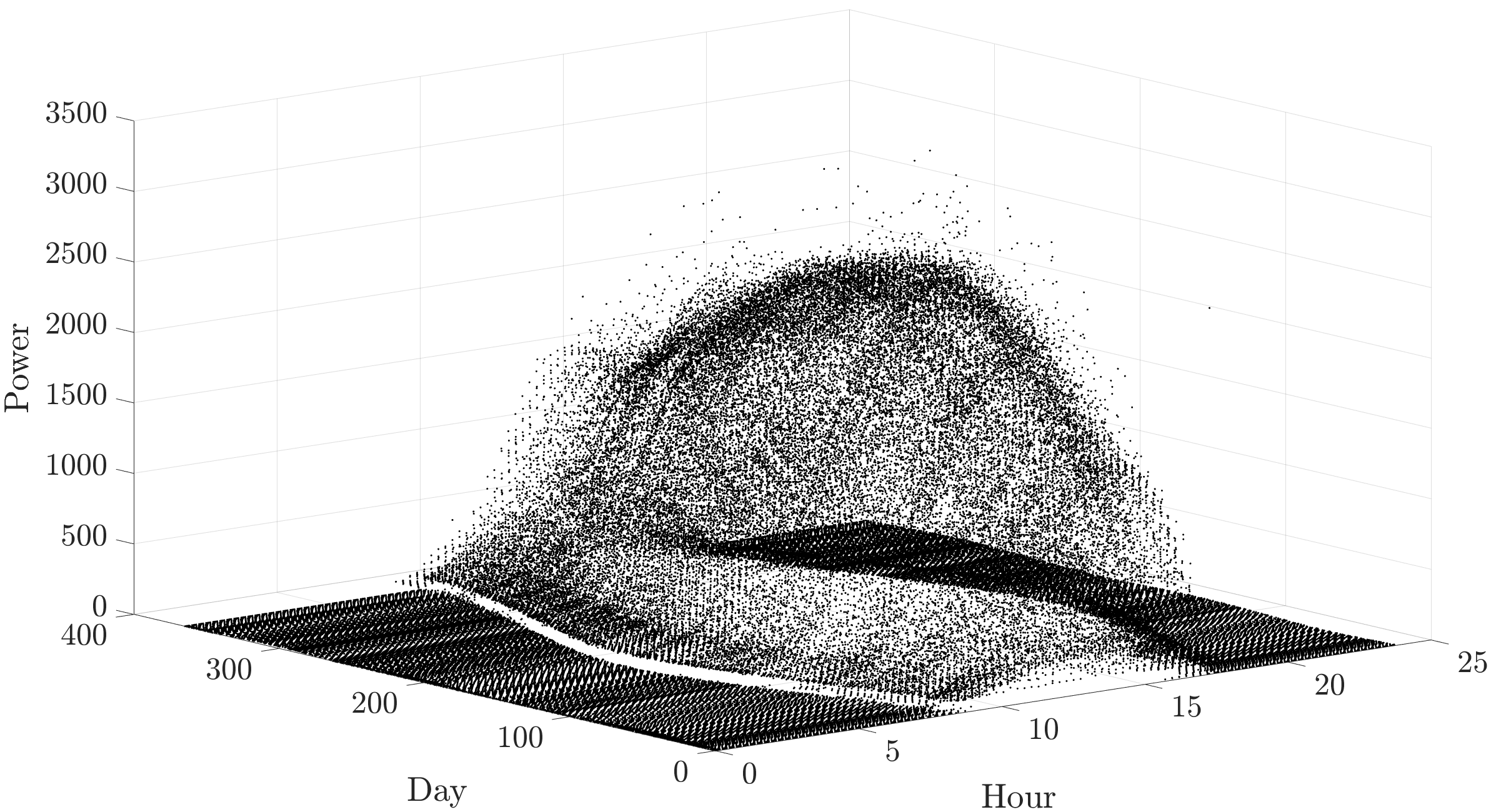}
	\caption{Plot of solar PV power with time and day of the year}
	\label{fig:3D_power_plot}
\end{figure}   	

	    	Fig. \ref{fig:3D_power_plot} shows the variability of power with time and day of the year at the installation in California. 
	    	
		Normalized mean square error (NMSE) was used as the error metric in the regression problem from section \ref{sec:Model}, 
		\begin{align}
		\text{NMSE}_{n} = \frac{\sum\limits_{k}(w_{n}[k] - \hat{w}_{n}[k])^2} {\sum\limits_{k}(w_{n}[k])^2} 
		\end{align}	 	
	As reported in \cite{Asilomar2016}, the maximum normalized mean square error (NMSE)  was approximately 0.05 which proved the good fit provided by the model. The efficacy of the regression problem motivated the stochastic models for all the three regimes. 
\par
\begin{figure}	
			\centering
			\includegraphics[width=\columnwidth,height=.22\textheight]{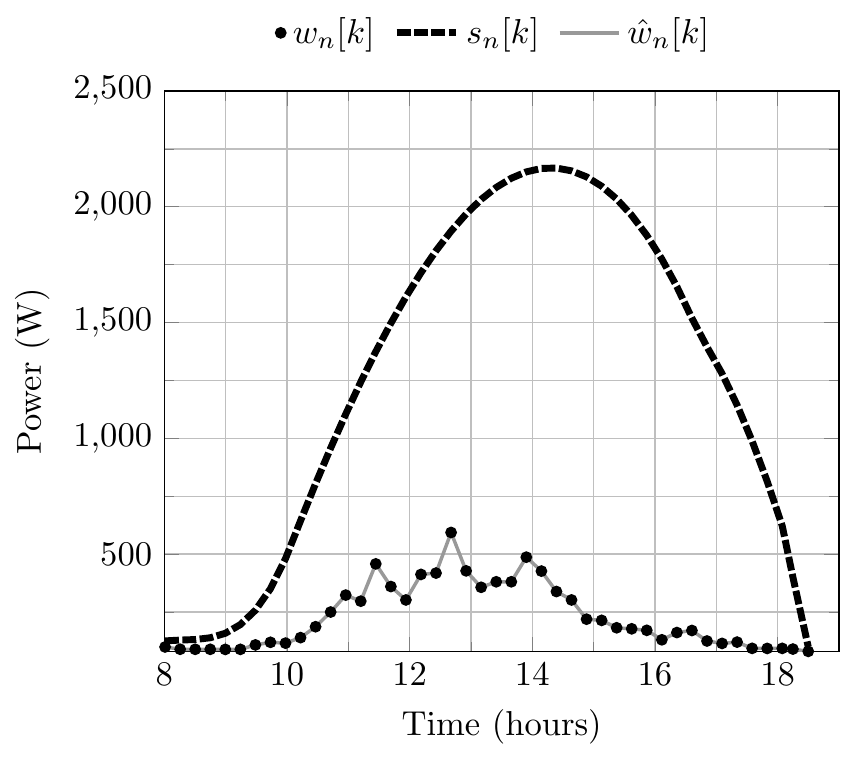}
			\caption{Fit of the model to an \textit{overcast day}}
			\label{fig:overcast_fit}
		\end{figure}

 As also seen in \cite{Asilomar2016}, the results for fitting highlight the switching of the solar PV power  between the three classes of \textit{sunny, overcast} and \textit{partly cloudy}. Furthermore, Fig.\ref{fig:overcast_fit} reflects the fitting for an overcast day with potentially persistent clouds that cause the power to look like a scaled version of the sunny day pattern. Therefore this observation was incorporated in the proposed model for the \textit{overcast} period in section \ref{sec:stochastic_models_for_classification}.
\par
\begin{figure}
	\centering
    \subfloat[Fit to a \textit{partly cloudy} day with sharp power fluctuations.\label{fig:sharp_fluct_fit}]{\includegraphics[width=\columnwidth,height=.22\textheight]{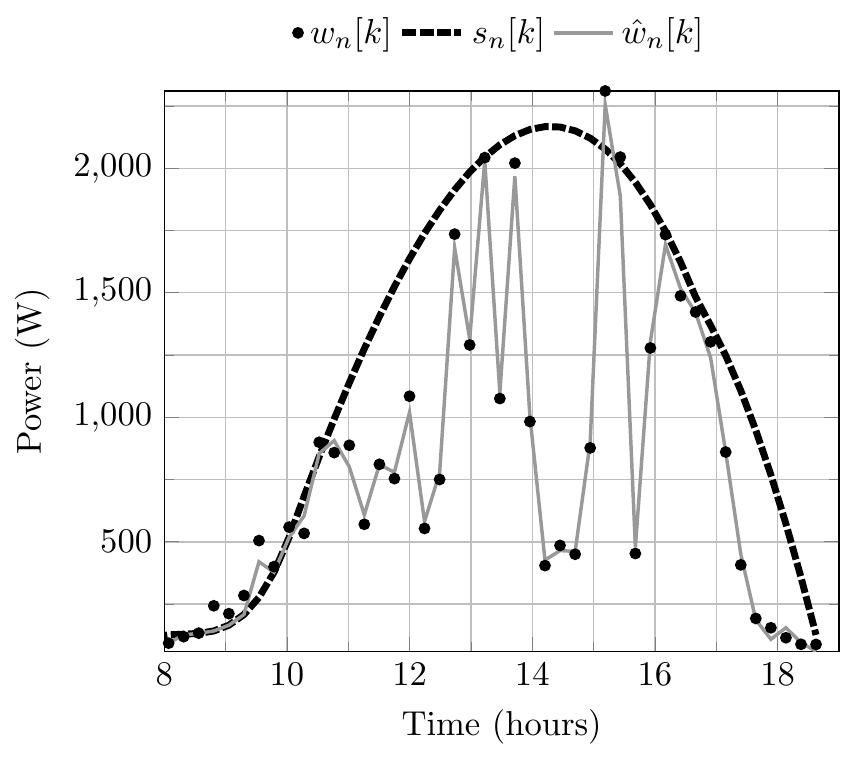}}\hfil
    \subfloat[Parameters for a \textit{partly cloudy} day with sharp power fluctuations \label{fig:sharp_fluct_params}]{\includegraphics[width=\columnwidth,height=.22\textheight]{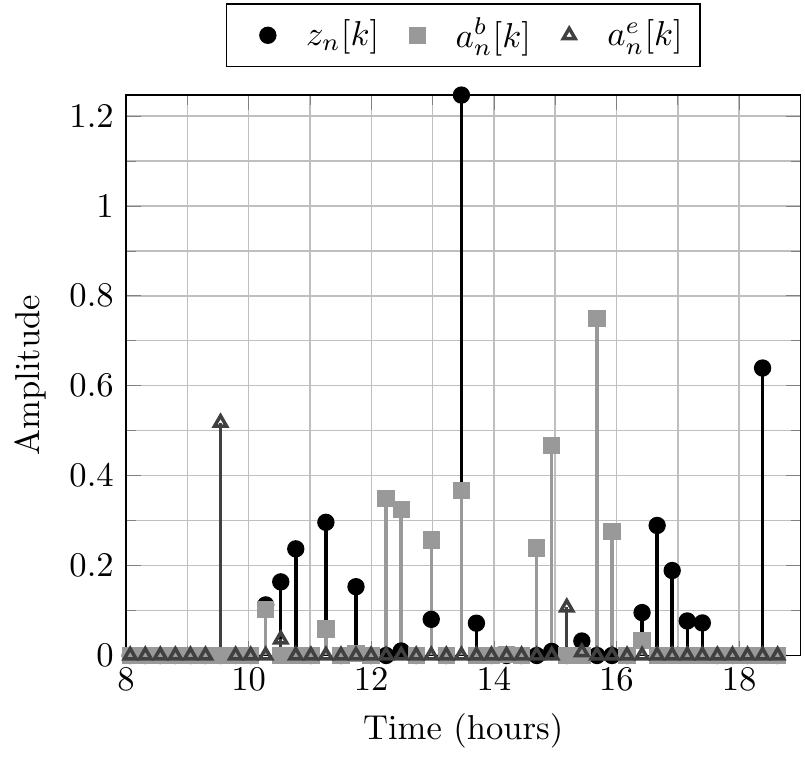}}\hfil
    \caption{Fitting a \textit{partly cloudy} day} \label{fig:partly_cloudy_total}
	\end{figure}

 Fig.\ref{fig:sharp_fluct_fit} highlights the fit of the model to a  \textit{partly cloudy} day. The parameters of $\mathbf{a}_{n}^{b}$ and $\mathbf{a}_{n}^{e}$ are larger and less sparse on such days, as shown in Fig.\ref{fig:sharp_fluct_params}. A natural result of inducing sparsity and non-negativity forced exponential distribution on the three parameters. Therefore, the \textit{partly cloudy} period was appropriately modeled as a HMM to capture the on-off process that characterizes these sparse components.
\subsection{Solar power prediction}\label{subsec:solar_power_pred_results}
In this subsection, the results of the prediction algorithm are provided.  Prediction horizon was $\chi =12$ i.e. $3$ hours and filter length, $M=5$ was used during \textit{partly cloudy} conditions. The algorithm started with $\chi_{window} =4$ samples ($1$ hour) for each day and predicted for the next $\chi$ samples. Then, the window was moved by one sample. 
\subsubsection{Metrics used for evaluation}
\par The results are presented using both deterministic and probabilistic forecast metrics. In the deterministic setting,
results are provided using the metrics of mean absolute percentage error (MAPE), $e_{abs}[{\kappa_{\tau}}]$ and root mean squared error (RMSE), $\text{RMSE}({k_{\tau}})$  for the $k_{\tau}$-step prediction ,
\begin{align}
e_{abs}[{\kappa_{\tau}}] &= \frac{ \sum\limits_{k,n} \frac{\lvert w_{n}[k]-\hat{w}^{k_{\tau}}_{n}[k] \rvert}{w_{n}[k]}} {\sum\limits_{k,n}1}\\
\text{RMSE}({k_{\tau}}) &= \sqrt{\frac{\sum\limits_{k,n} (w_{n}[k] - \hat{w}^{k_{\tau}}_{n}[k])^{2}}{\sum\limits_{k,n}1}}
\end{align}
where $\hat{w}^{k_{\tau}}_{n}[k]$ refers to the prediction at time $k$ given $w_{n}[k-k_{\tau}] $ and values before it.
\par In the probabilistic setting at each $k_{\tau}$-step prediction there is a cumulative distribution function (CDF) $F_{\hat{w}^{k_{\tau}}_{n}[k]}(x)$ instead of a point forecast $\hat{w}^{k_{\tau}}[k]$. Based on class of model chosen for prediction, the CDF is determined as 
\begin{align}
F_{\hat{w}^{k_{\tau}}_{n}[k]}(x) = \begin{cases}
\Phi(\frac{x - \hat{w}^{k_{\tau}}_{n}[k]}{\sigma_{s}}), \textit{sunny} \\
\int_{0}^{x} \tilde{f}_{oc}(x), \textit{overcast} \\
\int_{0}^{x} \tilde{f}_{i}(x),~ i=1,2,\dots,\mathcal{N}_{s}, \textit{partly cloudy}
\end{cases}
\end{align}
with $\Phi(.)$ denoting the CDF of a standard normal distribution, $\tilde{f}_{oc}(x)$ and $\tilde{f}_{i}(x)$ are defined as in \eqref{oc_PDF} and \eqref{cond_PDF} respectively.
The metrics used for evaluation are continuous rank probability score (CRPS) \cite{Gneiting2007}, reliability metric and score \cite{probabilistic_scores}. 
CRPS is defined for each $k_{\tau}$-step prediction as an average over all the samples,
\begin{align}
\text{CRPS}(k_{\tau})= \frac{\sum\limits_{k,n} \int_{0}^{\infty}\left(F_{\hat{w}^{k_{\tau}}_{n}[k]}(y) - u(y-w_{n}[k])\right)^{2} dy} {\sum\limits_{k,n}1}
\end{align} where $u(.)$ is the Heaviside step function.
The CRPS evaluates to mean squared error (MSE) when the forecast is deterministic. 
\par Reliability of a probabilistic forecasting method is a useful metric in understanding  the  proximity of the estimated CDF to the actual CDF of the data. Let a probability interval (PI), $I_{\hat{w}^{k_{\tau}}_{n}[k]}$, be defined with an upper and lower bound such that the interval covers the observed value $w_{n}[k]$ with probability $(1-\tilde{b})$. Then, to calculate reliability, define
\begin{align}
R_{\tilde{b}}(k_{\tau}) = \frac{\sum\limits_{k,n} \mathbb{I}_{\left(w_{n}[k] \in I_{\hat{w}^{k_{\tau}}_{n}[k]}\right)}}{\sum\limits_{k,n}1}
\end{align}
as the estimated probability of coverage where $\mathbb{I}_{\left(.\right)}$ is an indicator function with value $1$ if the observed sample belongs to the probability interval. Now, the probabilistic forecast is more reliable if the quantity $\tilde{R}_{\tilde{b}}$
\begin{align}
\tilde{R}_{\tilde{b}} (k_{\tau}) \triangleq R_{\tilde{b}} (k_{\tau})- (1-\tilde{b})
\end{align} is small. 
\par Another metric used for evaluation is the score. This metric is helpful in determining the sharpness of the forecast probability interval by imposing a penalty when an observation is outside the interval, by a value proportional to the size of the interval. If the upper and lower bounds of the PI are denoted as $\tilde{U}_{\hat{w}^{k_{\tau}}_{n}[k]}$ and $\tilde{L}_{\hat{w}^{k_{\tau}}_{n}[k]}$ respectively then score is defined as
\begin{align}
&\text{Score}_{\tilde{b}}^{k_{\tau}}[k] = \nonumber\\
&\begin{cases}
\tilde{D}_{\hat{w}^{k_{\tau}}_{n}[k]} - 4 \left(\tilde{L}_{\hat{w}^{k_{\tau}}_{n}[k]} - w_{n}[k]\right) , ~w_{n}[k] < \tilde{L}_{\hat{w}^{k_{\tau}}_{n}[k]}\\
\tilde{D}_{\hat{w}^{k_{\tau}}_{n}[k]} , ~ w_{n}[k] \in I_{\hat{w}^{k_{\tau}}_{n}[k]} \\
\tilde{D}_{\hat{w}^{k_{\tau}}_{n}[k]} - 4 \left(w_{n}[k] - \tilde{U}_{\hat{w}^{k_{\tau}}_{n}[k]} \right) , ~w_{n}[k] > \tilde{U}_{\hat{w}^{k_{\tau}}_{n}[k]}
\end{cases}
\end{align}
where 
\begin{align}
\tilde{D}_{\hat{w}^{k_{\tau}}_{n}[k]} \triangleq -2\tilde{b} \left(\tilde{U}_{\hat{w}^{k_{\tau}}_{n}[k]}-\tilde{L}_{\hat{w}^{k_{\tau}}_{n}[k]}\right)
\end{align}
The average score is,
\begin{align}
\text{Score}_{\tilde{b}}(k_{\tau}) = \frac{\sum\limits_{k,n}\text{Score}_{\tilde{b}}^{k_{\tau}}[k]}{\sum\limits_{k,n}1}
\end{align}
Lower values of the score indicate sharper and more reliable forecasts.
\par
\begin{figure}
	\centering
	\includegraphics[width=\columnwidth,height=.22\textheight]{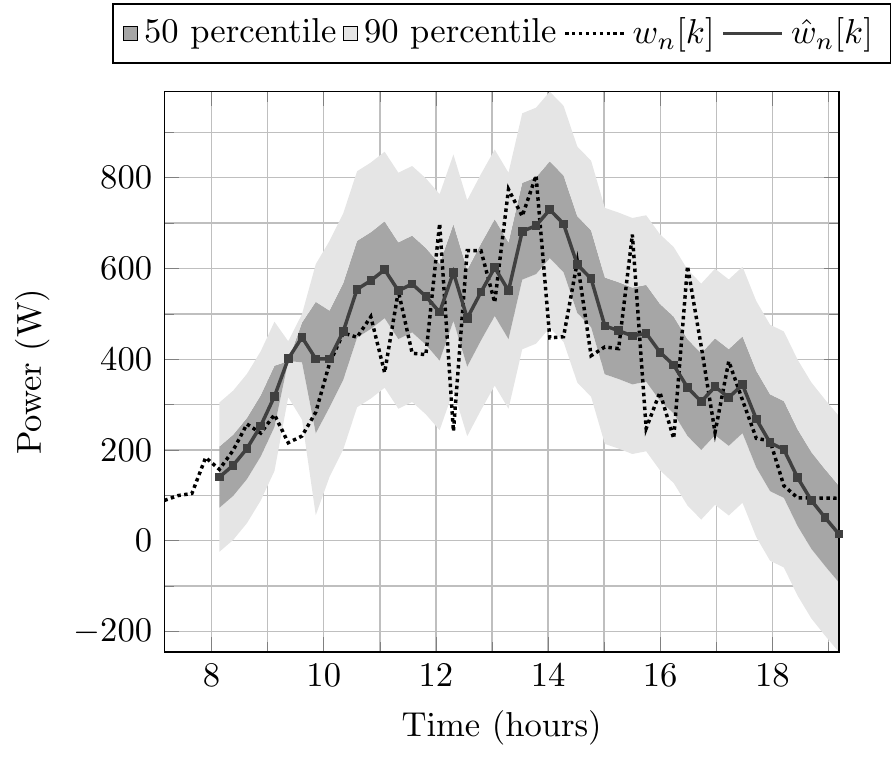}
	\caption{Plot of actual and predicted value with one-step prediction for day that is \textit{overcast}}
	\label{fig:one_step_pred_59}
\end{figure}

\begin{figure}
	\centering
	\includegraphics[width=\columnwidth,height=.22\textheight]{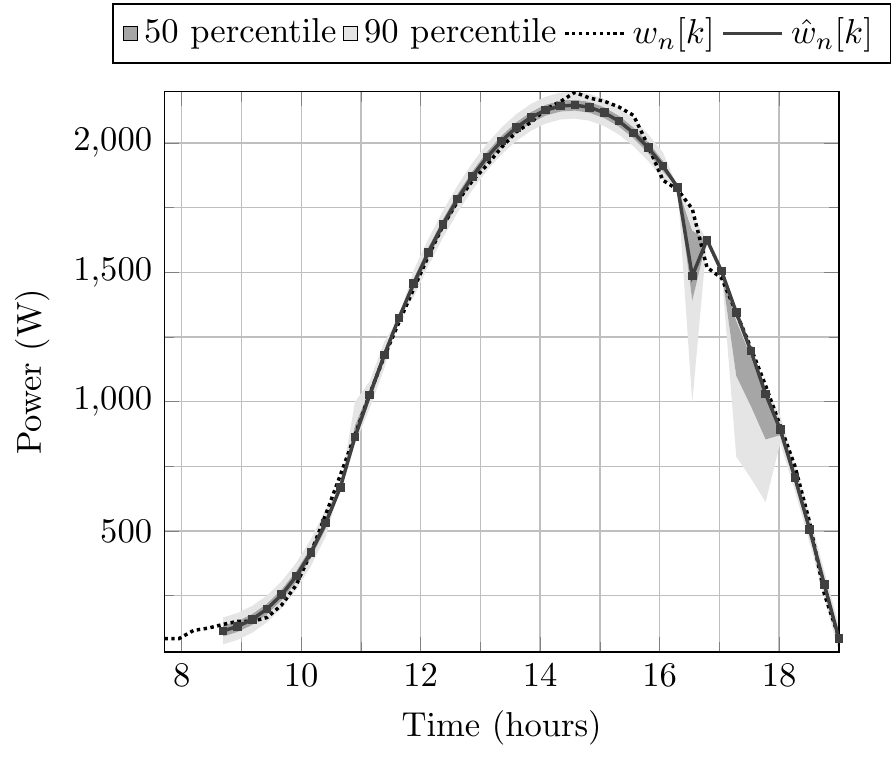}
	\caption{Plot of actual and predicted value with one-step prediction for day that is \textit{sunny} day}
	\label{fig:one_step_pred_65}
\end{figure}

\begin{figure}
	\centering
	\includegraphics[width=\columnwidth,height=.22\textheight]{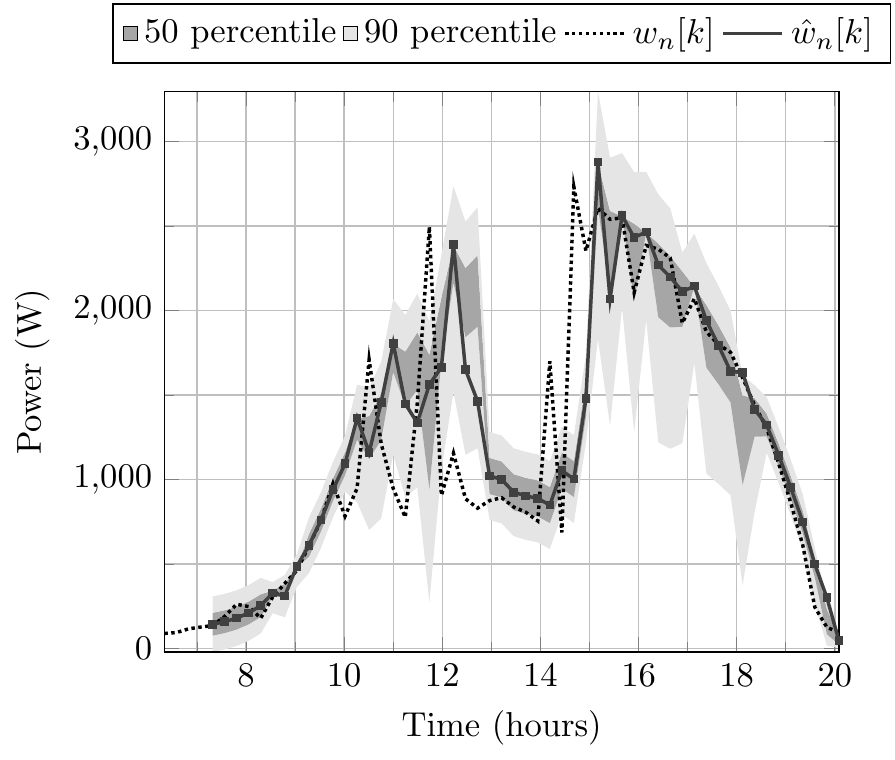}
	\caption{Plot of actual and predicted value with one-step prediction for day with variety of weather conditions}
	\label{fig:one_step_pred_55}
\end{figure}

 Performance of the prediction methods is analyzed using average reliability and score defined as
\begin{align}
{R}^{\text{avg}}_{\tilde{b}} &= \sum\limits_{k_{\tau}}R_{\tilde{b}} (k_{\tau})/ \chi\\
 \text{Score}^{\text{avg}}_{\tilde{b}} &= \sum\limits_{k_{\tau}}\text{Score}_{\tilde{b}} (k_{\tau})/ \chi
\end{align}
As representative examples, Fig. \ref{fig:one_step_pred_59}, \ref{fig:one_step_pred_65} and \ref{fig:one_step_pred_55}  show the actual and predicted power for different days with a multitude of weather conditions. This predicted power is one-step prediction, $k_{\tau}=1$. 
For days that are entirely \textit{overcast} or \textit{sunny}, predictions have little error as can be seen in Fig. \ref{fig:one_step_pred_59} and \ref{fig:one_step_pred_65}. These results highlight that the stochasticity of power in both these regimes is minimal leading to better predictions if the weather condition persists. 
 However, there is  higher error whenever there is a change in regime, for example going from \textit{partly cloudy} condition to \textit{overcast} around $12$ PM as seen in Fig. \ref{fig:one_step_pred_55}. It can be attributed to the delay in detecting the change in model. This uncertainty cannot be avoided in days with sudden change in weather unless there is some additional information in the form of cloud motion information or accurate weather forecasts. 
 To summarize, prediction during \textit{partly cloudy} conditions is prone to larger errors than during \textit{overcast} or \textit{sunny}. This is in accordance with the associated uncertainty in solar PV power for those periods. 
		\subsubsection{Comparison with reference models}\label{subsubsec:comaprison_pred}
The proposed prediction method is compared with reference methods  in both deterministic and probabilistic setting.
Since the proposed method does not use external inputs,  comparison is made with respect to multiple reference models that do not need any external input and are generally used as  benchmarks.
\par  The benchmarks considered for point forecast comparison are  diurnal persistence, smart persistence \cite{smart_persist},  artificial neural networks (ANN), AR model and AR model with regime switching. 
\par In \textit{diurnal persistence}, the prediction for a time $k$ is the power value at the same time on the previous day if available,
	\begin{align}
	\hat{w}^{diurnal}_{n}[k]=w_{n-1}[k], ~k = \kappa_{2}+1,\dots,\kappa_{2}+\chi \label{eq:diurnal}
	\end{align}
\par In smart persistence,  the prediction for the next $k$ steps is given as the continued fraction of clear sky component at the current time step,
\begin{align}
 \hat{w}^{per}_{n} [k] =  s_{n}[k] \frac{w_{n}[\kappa_{2}]}{s_{n}[\kappa_{2}]}, ~ k = \kappa_{2}+1,\dots,\kappa_{2}+\chi \label{eq:persistence}
\end{align}
\par In the AR model,  the power is expressed as the sum of clear sky component and stochastic component, and is assumed that the stochastic	component has an autoregressive model:
\begin{align}
w^{AR}_{n}[k] = s_{n}[k] + x^{AR}_{n}[k], \\
x^{AR}_{n}[k] = \sum\limits_{i=1}^{M^{AR}} a[i] x^{AR}_{n}[k-i] + \epsilon_{AR}[k]
\end{align} 
In AR model with regime switching, it is assumed that each of the classes \textit{sunny, partly cloudy} and \textit{overcast} have stochastic components with different coefficients corresponding to the AR model:
\begin{align}
x^{AR}_{n}[k] =\begin{cases}
\sum\limits_{i=1}^{M^{AR_s}} a_{s}[i] x^{AR}_{n}[k-i]+ \epsilon_{AR_{s}}[k], ~ \textit{sunny} \\
\sum\limits_{i=1}^{M^{AR_{pc}}} a_{pc}[i]  x^{AR}_{n}[k-i]+ \epsilon_{AR_{pc}}[k], ~\textit{partly cloudy}\\
 \sum\limits_{i=1}^{M^{AR_{oc}}} a_{oc}[i]  x^{AR}_{n}[k-i]+ \epsilon_{AR_{oc}}[k], ~\textit{overcast}
\end{cases} 
\end{align}
In addition, the proposed method is also compared with the artificial neural network (ANN) approach. Specifically, a non-linear autoregressive neural network (NARNET)\cite{ANN_NARNET} was used. These are essentially feed-forward networks with autoregressive nature:
\begin{align}
w^{ANN}_{n}[k] &= s_{n}[k] + x^{ANN}_{n}[k],\\
{x}^{ANN}_{n}[k] &= \sum\limits_{i=1}^{L}W_{i}\sum\limits_{j=1}^{P}f\left(\tilde{\beta}_{ij}x^{ANN}_{n}[k-j] + \theta_{i}\right)
\end{align} 
$2$ hidden layers with $10$ neurons each and a lag $p=15$ was used for the stochastic component, $x^{ANN}_{n}[k]$. The activation function $f(.)$ was $tanh(x)= 2/(1+\exp{(-2x)}) - 1$. 
\par For comparison in the probabilistic forecast setting, smart persistence and AR models are used.  In the smart persistence approach, it is assumed that the smart persistence forecast in \eqref{eq:persistence} is the mean and the variance is estimated from the samples used for forecasting. The distribution is assumed to be Gaussian. 
\par In both the AR model and the regime switching AR models, the point forecast value is the mean and the variance of Gaussian noise, $\epsilon_{AR}$  is estimated along with the coefficients.
\subsubsection{Results}
           All the simulations were performed using one year of training data and one year of testing data for validation. The programs were written using MATLAB and executed on a machine with Intel i7 processor with 8GB RAM and at 2.2 GHz. Most of the training for estimation of parameters of HMM is done apriori making computational time of the proposed method very short since the Viterbi algorithm, which is proven to be efficient \cite{viterbi_algorithm} was used. The computational time specifically depends on the acquisition time of samples in a real-time setting. In the simulation, since data was already available, it took $4$ milliseconds on an average to make predictions for a horizon of $3$ hours at $15$ minute intervals, i.e. for $12$ samples ahead.
		  	\par
\begin{figure}
		  			\centering
		  			\includegraphics[width=\columnwidth,height=0.22\textheight]{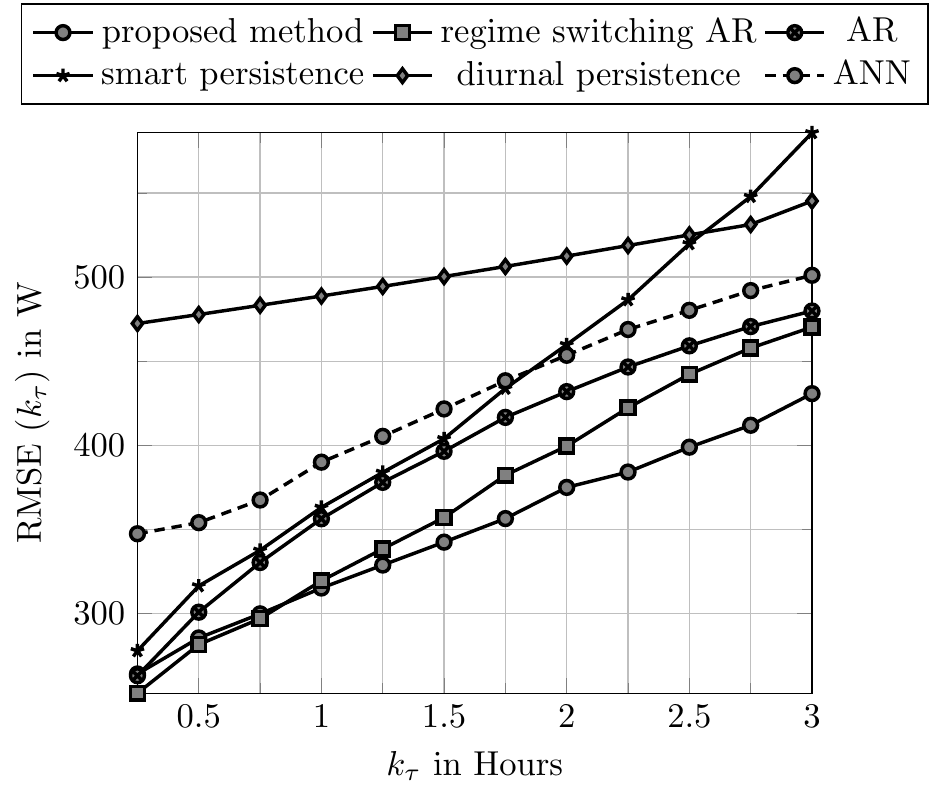}
		  			\caption{ RMSE for $k_{\tau}$-step prediction using proposed model and other reference models.}
		  			\label{fig:RMSE_pred}		  		
		  		\end{figure}

\begin{figure}
		  				\centering
		  				\includegraphics[width=\columnwidth,height=0.22\textheight]{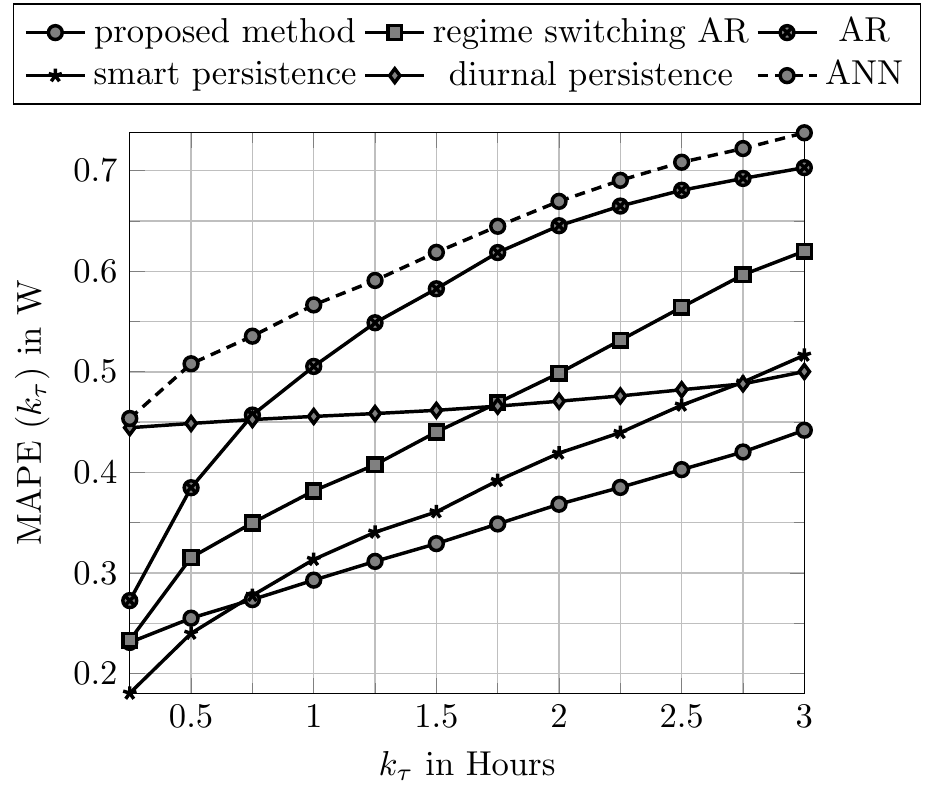}
		  				\caption{ MAPE for $k_{\tau}$-step prediction using proposed model and other reference models.}
		  				\label{fig:MAE_pred}
		  		\end{figure}

 Figs. \ref{fig:RMSE_pred} and \ref{fig:MAE_pred} depict the RMSE and MAPE respectively for various prediction horizons and compares different methods. 
		  	\par As seen from Fig. \ref{fig:RMSE_pred}, it is evident that the proposed method consistently outperforms the other methods used as benchmarks. The performance of all the methods is comparable when the prediction step is less than $30$ minutes. But as the horizon increases, diurnal persistence fares the worst with smart persistence and ANN coming close. ANN method fares badly with respect to MAPE as shown in Fig. \ref{fig:MAE_pred} indicating that architectural changes could be necessary with possibly more number of neurons in order to achieve better results. This means that more parameters and tuning is needed in ANNs whereas in comparison the proposed method uses fewer parameters and a low order model for prediction. 
		  	\par Both the types of AR models perform better than the naive benchmarks as reported by other works. The regime switching AR model approach is similar in performance to the proposed method indicating that regime switching is the appropriate type of method to use in the case of prediction for solar PV power data. 
		  	\par
\begin{figure}
		  			\centering
		  			\includegraphics[width=\columnwidth,height=0.23\textheight]{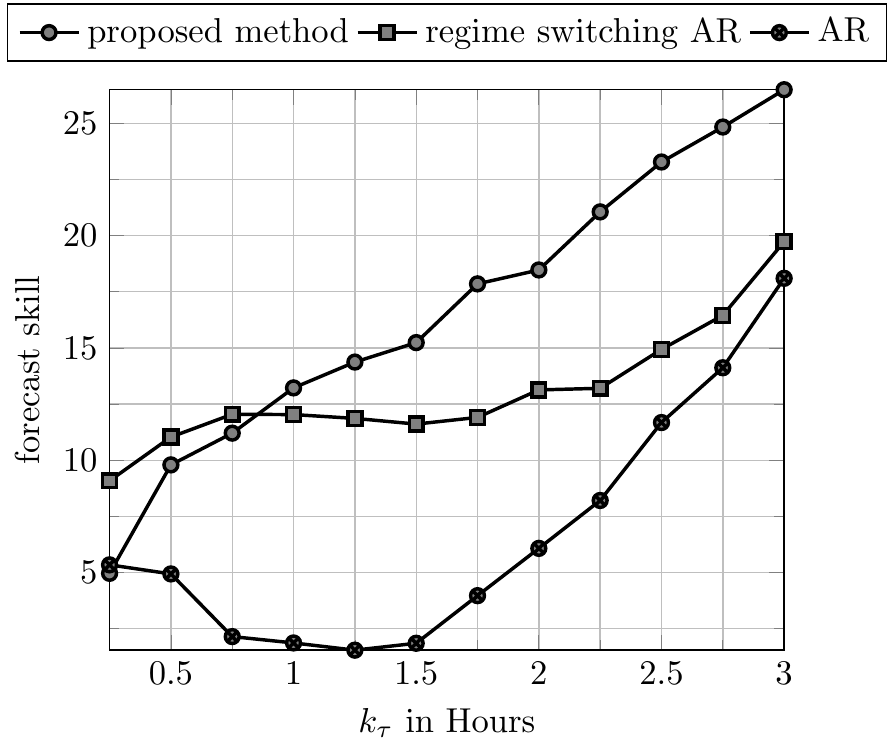}
		  			\caption{ Forecasting skill for  $k_{\tau}$-step prediction using proposed model and other reference models as a percentage of improvement over smart persistence.}
		  			\label{fig:f_skill}
		  		\end{figure}

 Fig. \ref{fig:f_skill} shows the forecasting skill \cite{Coimbra2016} which highlights the improvement in forecasting as compared to `smart persistence'. It is defined as,
		  	\begin{align}
		  	f_{skill} (\kappa_{\tau})= 1 - \frac{RMSE(\kappa_{\tau})}{RMSE_{per}(\kappa_{\tau})}
		  	\end{align} 
		  	The improvement of the proposed method over smart persistence increases to $20\%$ along with the prediction horizon.
		  	\par
\begin{figure}
		  					\centering
		  					\includegraphics[width=\columnwidth,height=0.2\textheight]{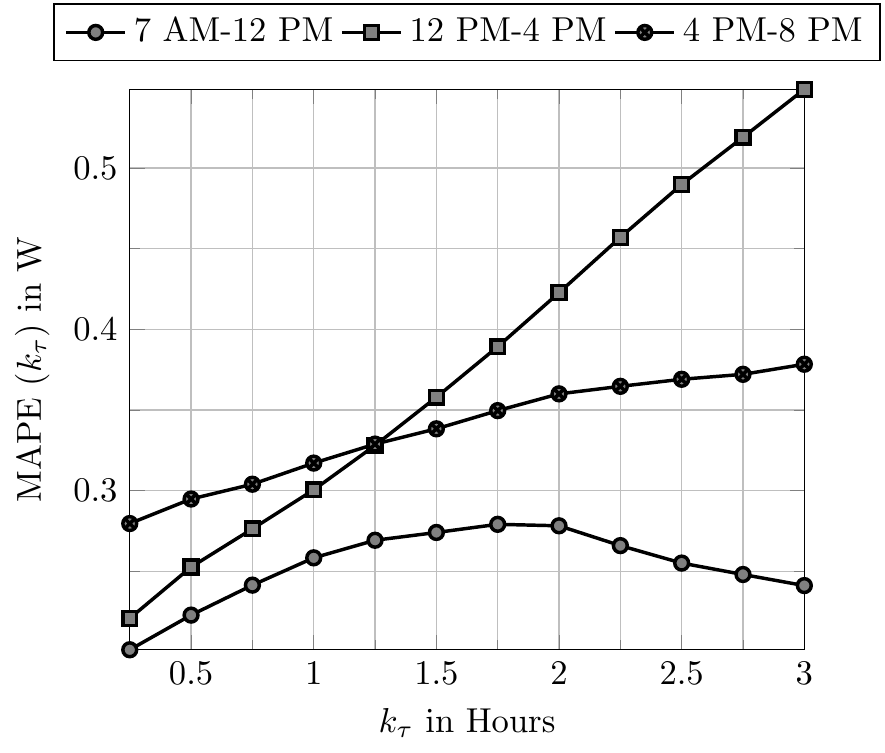}
		  					\caption{ MAPE for $k_{\tau}$-step prediction at different times during a day using proposed model.}
		  					\label{fig:MAPE_timing}
		  			\end{figure}

 Fig.\ref{fig:MAPE_timing} shows the MAPE averaged at different time intervals for prediction using the proposed method. As expected, the error is higher during the middle of the day when the uncertainty is very high. At morning and evening times of the day, lower variability in solar PV power improves prediction accuracy. This trend is obvious for other reference methods as well.
		  	\par A note to make here is that the values of $\lambda_{z}, \lambda_{a}^{b}, \lambda_{a}^{e}$ play a role in deciding the performance of the proposed methods. As the values of $\lambda_{z}, \lambda_{a}^{b}, \lambda_{a}^{e}$ increase, the point forecasting performance improves to a certain extent. 
		  	\par
\begin{figure}
		  		\centering
		  		\includegraphics[width=\columnwidth,height=0.25\textheight]{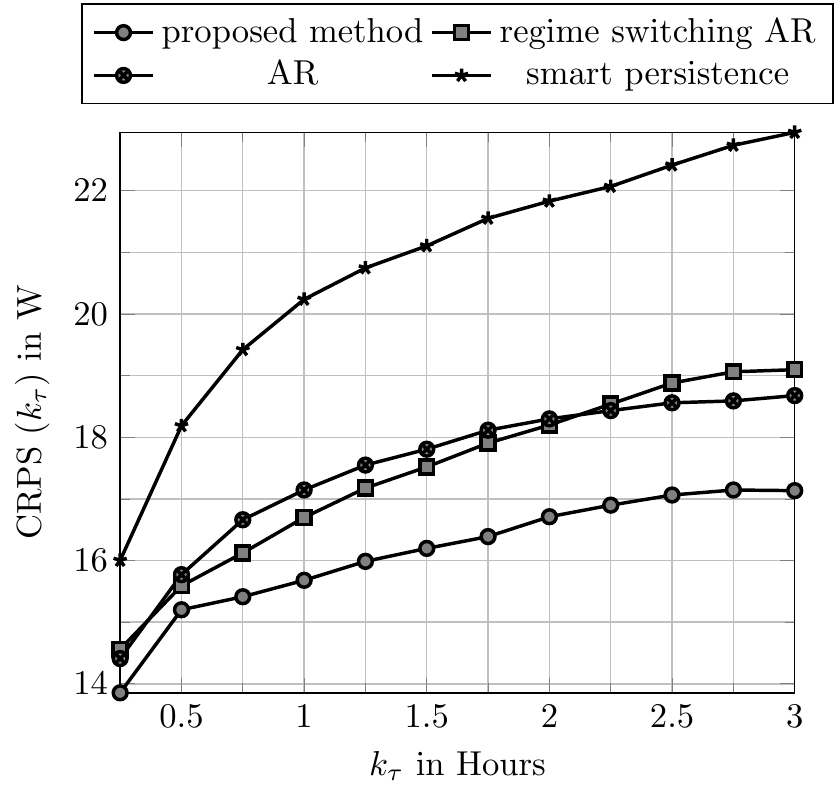}
		  		\caption{ CRPS for $k_{\tau}$-step prediction using proposed model and other reference models.}
		  		\label{fig:CRPS_Pred}
		  	\end{figure}

 Comparison with CRPS as the metric is shown in Fig.\ref{fig:CRPS_Pred}. Again, the trend among the competing methods is repeated from the deterministic setting with AR models beating the smart persistence. 
		  	\par
\begin{figure}
		  			\centering
		  			\includegraphics[width=\columnwidth,height=0.23\textheight]{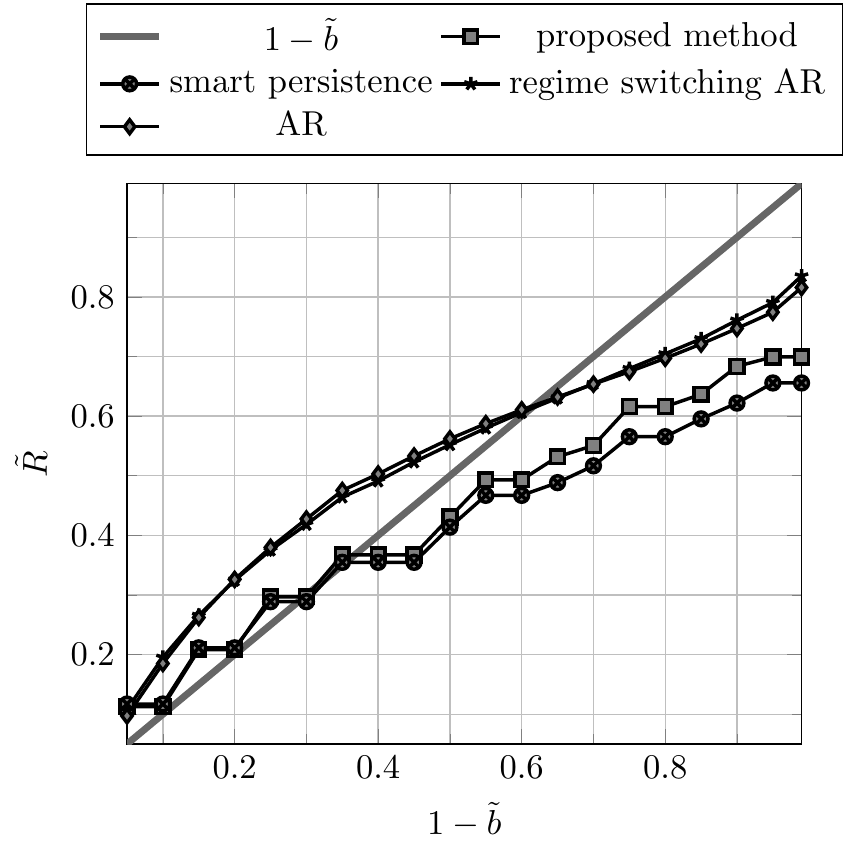}
		  			\caption{ Average reliability using proposed model and other reference models.}
		  			\label{fig:reliability}
		  		\end{figure}

 Fig.\ref{fig:reliability} shows the average reliability metric,${R}^{\text{avg}}_{\tilde{b}}$  and comparison with the curve $1-\tilde{b}$. The closer it is to the curve, the more reliable the method. From the figure, one can observe that the proposed method does well in the lower quantiles but not so well in the higher quantiles. This is due to the choice of the values of $\lambda_{z}, \lambda_{a}^{b}, \lambda_{a}^{e}$ being small. 
		  	\par Using the same set of parameters, Table.\ref{table:score} shows the average score, ${\text{Score}}^{\text{avg}}_{\tilde{b}}$ normalized by the nameplate capacity. In terms of score, the proposed method outperforms all the other methods considered. This is indicative of the fact that the forecasts from the proposed method are sharp and well calibrated in general.
		  \par 	It is pertinent to discuss  that  $\lambda_{z}, \lambda_{a}^{b}, \lambda_{a}^{e}$ and $\sigma_{s}, \sigma_{oc}$ are tuning parameters which affect the performance of the proposed method. Decreasing the values makes the predictions less reliable and sharp but fares well when seen from the CRPS perspective.  This is because CRPS only accounts for how well the forecast probability intervals cover the observed value of power which means that larger the intervals (smaller $\lambda_{z}, \lambda_{a}^{b}, \lambda_{a}^{e}$), better is the CRPS. But the metrics of score and reliability penalize wide intervals and therefore the method fares better with smaller width of intervals (larger $\lambda_{z}, \lambda_{a}^{b}, \lambda_{a}^{e}$ ). 
		  	\section{Discussion}\label{sec:Discussion}
		  	The proposed method is better suited for shorter horizons i.e. less than $4$ hours since persistence in weather condition is assumed. In the situation that no weather forecasts or other additional information is used, the performance of the proposed prediction algorithm is good and outperforms multiple benchmark models. It was concluded that the  regime switching AR model is closest in performance to the proposed method which shows the advantages of considering a regime switching approach since solar PV power data is non-stationary.
		  	\par More importantly,  the proposed model is stochastic and provides probabilistic forecasts of power over the desired horizon. The performance of the proposed method can be adjusted by tuning the parameters $\lambda_{z}, \lambda_{a}^{b}, \lambda_{a}^{e}$. Larger probability intervals generated with smaller values of the parameters are more suited  when the evaluation metric is CRPS. Narrower intervals are desirable for sharper and reliable forecasts. Based on the demand of the application at hand, the forecasts can be suitably adapted. 
		  	\par Sample future power scenarios can be produced by considering all three stochastic models to be probable in the future. These scenarios are quite useful while solving stochastic optimization problems such as designing a battery storage policy \cite{tutorial_stochastic_opt}. This is future work.

		  \begin{table}[h]
		  	\centering
		  	 \caption{Table with average score  normalized by the nameplate capacity}
		  \begin{tabular}{|l|c|c|c|c|}%
		  \hline 
		  	\bfseries $1-\tilde{b}$ & \bfseries Proposed & \bfseries AR with switching &\bfseries Persistence & \bfseries AR   
		  	\begin{filecontents*}{Score_all_methods_normalized.csv}
alpha,S,Sper,SAR,SARSwitch,
0.05,-0.218,-0.241,-0.249,-0.221,
0.10,-0.219,-0.242,-0.239,-0.212,
0.15,-0.207,-0.230,-0.232,-0.205,
0.20,-0.208,-0.232,-0.227,-0.201,
0.25,-0.200,-0.225,-0.224,-0.198,
0.30,-0.203,-0.228,-0.224,-0.198,
0.35,-0.199,-0.226,-0.224,-0.200,
0.40,-0.203,-0.231,-0.228,-0.204,
0.45,-0.207,-0.235,-0.233,-0.209,
0.50,-0.209,-0.240,-0.240,-0.218,
0.55,-0.214,-0.249,-0.249,-0.228,
0.60,-0.221,-0.256,-0.261,-0.242,
0.65,-0.228,-0.267,-0.275,-0.257,
0.70,-0.241,-0.282,-0.293,-0.277,
0.75,-0.263,-0.309,-0.313,-0.300,
0.80,-0.273,-0.320,-0.339,-0.330,
0.85,-0.299,-0.351,-0.373,-0.367,
0.90,-0.330,-0.383,-0.415,-0.414,
0.95,-0.381,-0.445,-0.468,-0.473,
0.99,-0.393,-0.459,-0.558,-0.575,
\end{filecontents*}
\csvreader[head to column names]{Score_all_methods_normalized.csv}{}
		  	{\\\hline \alpha& \S & \SARSwitch & \Sper & \SAR }
		  	  \\\hline
		  \end{tabular}				 
		  \label{table:score}
		  \end{table}
		 
	\section{Conclusions}\label{sec:Conclusions}
		A regime-switching process was proposed for the depiction and prediction of solar PV power. Stochastic models for different periods of \textit{sunny, overcast} and \textit{partly cloudy} were proposed along with an online, computationally efficient algorithm for short term probabilistic forecasts. The prediction algorithm was shown to compare favorably with many reference models. It was also shown that the prediction algorithm is tunable and depending on the end goal, one can suitably adapt the performance. 
Future work includes accounting for the spatial correlation in solar power at multiple locations through low order models and extending the model to provide probabilistic forecasts at different locations simultaneously. 
\vspace{-.2cm}	\bibliographystyle{IEEEtran}
	\bibliography{IEEEabrv,references_solar}
\end{document}